\documentclass[usenatbib]{mn2e}
\usepackage{amssymb,amsmath,graphicx,natbib,setspace,pifont}
\newcommand{\Hm}{\rm H_{2} }
\newcommand{\NHI}{{N_{\rm HI}}}
\newcommand{\fNHI}{f(N_{\rm HI},z)}

\newcommand{\cmsq}{\,{\rm cm^{-2}}}
\newcommand{\cms}{\,{\rm cm^{2}}}
\newcommand{\cmcb}{\,{\rm cm^{-3}}}

\newcommand{\nH}{n_{\rm _{H}} }

\newcommand{\nHI}{n_{\rm _{HI}}}

\newcommand{\Ob}{\Omega_{\rm b} }
\newcommand{\Om}{\Omega_{\rm m} }

\newcommand{\Ol}{\Omega_{\Lambda} }
\newcommand{\ns}{n_{\rm s} }
\newcommand{\sigeight}{\sigma_{\rm 8} }

\newcommand{\Msun}{{\rm M_{\odot}} }
\newcommand{\Msunh}{h^{-1} {\rm M_{\odot}} }
\newcommand{\Mpch}{h^{-1} {\rm Mpc} }
\newcommand{\kpch}{h^{-1} {\rm kpc} }

\newcommand{\Gadget}{{\small GADGET-3} }

\newcommand{\TRAPHIC}{{\small TRAPHIC}~}

\newcommand{\apjl}{{ApJL} }
\newcommand{\apj}{{ApJ} }
\newcommand{\apjs}{{ApJS} }
\newcommand{\mnras}{{MNRAS} }

\begin{document}
\title[The impact of local stellar radiation on the HI CDDF]{The impact of local stellar radiation on the HI column density distribution}

\author[A.~Rahmati et al.]
  {Alireza~Rahmati$^1$\thanks{rahmati@strw.leidenuniv.nl}, Joop Schaye$^1$, Andreas H. Pawlik$^2$, Milan Rai\v{c}evi\`{c}$^1$\\
  $^1$Leiden Observatory, Leiden University, P.O. Box 9513, 2300 RA Leiden, The Netherlands\\
  $^2$Max-Planck Institute for Astrophysics, Karl-Schwarzschild-Strasse 1, 85748 Garching, Germany}

\maketitle

\begin{abstract}
It is often assumed that local sources of ionizing radiation have little impact on the distribution of neutral hydrogen in the post-reionization Universe. While this is a good assumption for the intergalactic medium, analytic arguments suggest that local sources may typically be more important than the meta-galactic background radiation for high column density absorbers ($\NHI > 10^{17} \cmsq$). We post-process cosmological, hydrodynamical simulations with accurate radiation transport to investigate the impact of local stellar sources on the column density distribution function of neutral hydrogen. 
We demonstrate that the limited numerical resolution and the simplified treatment of the interstellar medium (ISM) that are typical of the current generation of cosmological simulations provide significant challenges, but that many of the problems can be overcome by taking two steps. First, using ISM particles rather than stellar particles as sources results in a much better sampling of the source distribution, effectively mimicking higher-resolution simulations. Second, by rescaling the source luminosities so that the amount of radiation escaping into the intergalactic medium agrees with that required to produce the observed background radiation, many of the results become insensitive to errors in the predicted fraction of the radiation that escapes the immediate vicinity of the sources. By adopting this strategy and by drastically varying the assumptions about the structure of the unresolved ISM, we conclude that we can robustly estimate the effect of local sources for column densities $\NHI \ll 10^{21} \cmsq$. However, neither the escape fraction of ionizing radiation nor the effect of local sources on the abundance of $\NHI \gtrsim 10^{21}\cmsq$ systems can be predicted with confidence. We find that local stellar radiation is unimportant for $\NHI \ll 10^{17} \cmsq$, but that it can affect Lyman Limit and Damped Ly$\alpha$ systems. For $10^{18} < \NHI < 10^{21} \cmsq$ the impact of local sources increases with redshift. At $z=5$ the abundance of absorbers is substantially reduced for $\NHI \gg 10^{17} \cmsq$, but at $z=0$ the effect only becomes significant for $\NHI \gtrsim 10^{21}\cmsq$. 
\end{abstract}


\section{Introduction}
After the reionization of the Universe at $z \gtrsim 6$, hydrogen residing in the intergalactic medium (IGM) is kept highly ionized primarily by the meta-galactic UV background (UVB). The UVB is the integrated radiation that has been able to escape from sources into the IGM. Because the mean free path of ionizing photons is large compared to the scale at which the sources of ionizing radiation cluster, the UVB is expected to be close to uniform in the IGM. However, close to galaxies, the radiation field is dominated by local sources and hence more inhomogeneous. 

Observations of neutral hydrogen (HI) in the Ly$\alpha$ forest mostly probe the low-density IGM which is typically far from star-forming regions. The statistical properties of the Ly$\alpha$ forest are therefore insensitive to the small-scale fluctuations in the UVB \citep[e.g.,][]{Zuo92,Croft04}. On the other hand, the neutral hydrogen in Damped Ly$\alpha$ (DLA; i.e., $\NHI > 10^{20.3}\cmsq$) and Lyman Limit systems (LL; i.e., $10^{17.2} <\NHI \leq 10^{20.3}\cmsq$), which are thought to originate inside or close to galaxies, might be substantially affected by radiation from local sources that are stronger than the ambient UVB \citep{Gnedin10}. As a result, the abundances of the high HI column densities may also change significantly by locally produced radiation. 

Indeed, \citet{Schaye06} and \citet{Miralda05} have used analytic arguments to show that the impact of local radiation may be substantial for LL and (sub-)DLA systems, but should generally be very small at lower column densities. However, relatively little has been done to go beyond idealized analytic arguments and to simulate the effect of local radiation by taking into account the inhomogeneous distribution of sources and gas in and around galaxies. One of the main reasons for this is the computational expense of radiative transfer (RT) calculations in simulations with large numbers of sources. In addition, high resolution is required to capture the distribution of gas on small scales accurately. Because of these difficulties, most simulations of the cosmological HI distribution have ignored the impact of local radiation and focused only on the effect of the UVB \citep[e.g.,][]{Katz96, Gardner97, Haehnelt98,Cen03,Nagamine04,Razoumov06, Pontzen08,Altay11,McQuinn11,Voort12,Rahmati12,Bird13}. A few studies have taken into account local stellar radiation but their results are inconclusive: while \citet{Nagamine10} and \citet{Yajima12} found that local stellar radiation has a negligible impact on the distribution of HI, \citet{Fumagalli11} found that the HI column density distribution above the Lyman limit is reduced by $\sim 0.5$ dex due to local stellar radiation.

In this paper, we investigate the impact of local stellar radiation on the HI distribution by combining cosmological hydrodynamical simulations with accurate RT. We find that the inclusion of local sources can dramatically change the predicted abundance of strong DLA systems and, depending on redshift, also those of LL and weak DLA systems. On the other hand, lower HI column densities are hardly affected. We also show that resolution effects have a major impact. For instance, the resolution accessible to current cosmological simulations is insufficient to resolve the interstellar medium (ISM) on the scales relevant for the propagation of ionizing photons. The limited resolution also affects the source distribution which can change the resulting HI distribution, especially in low-mass galaxies. On top of that, assumptions about the structure of the unresolved multiphase ISM significantly affect the escape of stellar radiation into the IGM. Therefore, any attempt to use cosmological simulations to investigate the impact of local stellar radiation on the HI distribution may suffer from serious numerical artifacts.

Some of these difficulties can be circumvented by tuning the luminosities of the sources such that the escaped radiation can account for the observed UVB. Then the interaction between the radiation that reaches the IGM and the intervening gas can be studied on scales that are properly resolved in the simulations. We adopt this procedure to generate the observed UVB for various ISM models but find that our fiducial simulation reproduces the observed UVB without any tuning.

Among the known sources of radiation, quasars and massive stars are the most efficient producers of hydrogen ionizing photons and are therefore thought to be the main contributors to the UVB. Star-forming galaxies however are thought to be the dominant producers of the UVB at $z \gtrsim 3$ \citep[e.g.,][]{Haehnelt01,Bolton05,Faucher08}. Therefore, we only account for local radiation that is produced by star formation in our simulations. Moreover, we assume that the ionizing emissivity of baryons strictly follows the star formation rate. We use star-forming gas particles rather than young stellar particles as ionizing sources. Since the gas consumption time scale in the ISM is much larger than the lifetime of massive stars ($\sim10^{9}$ yr vs. $\sim10^{7}$ yr), there are many more star-forming gas particles than there are stellar particles young enough to efficiently emit ionizing radiation. Therefore, using star-forming gas particles as ionizing sources allows us to sample the source distribution better and hence to reduce the impact of the limited resolution of cosmological simulations.

The structure of the paper is as follows. in $\S$\ref{sec:Gamma-KS} we use the observed relation between the star formation rate and gas surface densities to provide an analytic estimate for the photoionization rate that is produced by young stars in a uniform and optically thick ISM that is in good agreement with our simulation results. In $\S$\ref{sec:ingredients} we discuss the details of our numerical simulations and RT calculations. The simulation results are presented in $\S$\ref{sec:results}. We show how the local stellar radiation can generate the observed UVB. We also discuss the effect of unresolved ISM on the escape fraction of ionizing radiation and we investigate the impact of local stellar radiation on the HI column density distribution. Finally, we conclude  in $\S$\ref{sec:conclusions}.

\section{Photoionization rate in star-forming regions}
\label{sec:Gamma-KS}
A strong correlation between gas surface density and star formation rate has been observed in low-redshift galaxies \citep[e.g.,][]{Kennicutt98,Bigiel08}. In principle, such a relation can be combined with simplified assumptions to derive a typical photoionization rate that is expected from stellar radiation. In this section, we use this approach to estimate the average ionization rate that is produced by star formation as a function of gas (surface) density on galactic scales.

Assuming a \citet{Chabrier03} initial mass function (IMF), the observed Kennicutt-Schmidt law provides a relation between gas surface density, $\Sigma_{\rm{gas}}$, and star formation rate surface density, $\dot{\Sigma}_{\star} $, on kilo-parsec scales \citep{Kennicutt98}:
\begin{equation}
\dot{\Sigma}_{\star} \approx 1.5 \times 10^{-4}~ \Msun{\rm{yr^{-1}kpc^{-2}}} \left(\frac{\Sigma_{\rm{gas}}}{1\Msun{\rm{pc^{-2}}}} \right)^{1.4}.
\label{eq:KS-law}
\end{equation}
Equation \eqref{eq:KS-law} can be used to derive a relation between ionizing emissivity (per unit area), $\dot{\Sigma}_{\gamma}$, and the gas surface density. As we will discuss in $\S$\ref{sec:locals-def}, stellar population synthesis models indicate that the typical number of ionizing photons produced per unit time by a constant star formation rate is 
\begin{equation}
\dot{\mathcal{Q}}_{\gamma} \sim 2\times 10^{53}~\rm{s^{-1}}~\left(\frac{\rm{SFR}}{1~\Msun~{\rm{yr}}^{-1}}\right).
\label{eq:N-gamma_SFR}
\end{equation}
Furthermore, as both models and observations suggest, the escape fraction of ionizing photons from galaxies is $\ll1$ \citep[e.g.,][]{Shapley06,Gnedin08,Vanzella10,Yajima12,Paardekooper11,Kim12}. This allows us to assume that most of the ionizing photons that are produced by star-forming gas are absorbed on scales $\lesssim$ kpc. Therefore, the hydrogen photoionization rate, on galactic scales, can be computed:
\begin{equation}
\Gamma_{\star} = \frac{\dot{\Sigma}_{\gamma}}{N_{\rm H} }= \frac{\dot{\mathcal{Q}}_{\gamma}~ \dot{\Sigma}_{\star}}{N_{\rm H} },
\label{eq:Gamma-Sigma}
\end{equation}
where $N_{\rm H}$ is the hydrogen column density which can be obtained from the gas surface density: 
\begin{equation}
N_{\rm H} \approx 9.4 \times 10^{19}\cmsq~ \left(\frac{\Sigma_{\rm{gas}}}{1\Msun{\rm{pc^{-2}}}} \right)~\left(\frac{X}{0.75} \right),
\label{eq:NH-Sigma}
\end{equation}
where $X$ is the hydrogen mass fraction. After assuming $X = 0.75$ and substituting equations \eqref{eq:KS-law}, \eqref{eq:N-gamma_SFR} and \eqref{eq:NH-Sigma} in equation \eqref{eq:Gamma-Sigma}, one gets
\begin{equation}
\Gamma_{\star} \sim 8.5 \times 10^{-14}~{\rm{s^{-1}}}\left(\frac{N_{\rm H}}{10^{21}\cmsq} \right)^{0.4}.
\label{eq:Gamma-gasSigma}
\end{equation}
If the scale height of the disk is similar to the local Jeans scale, the hydrogen column density can be computed as a function of the hydrogen number density \citep{Schaye01,Schaye04,Schaye08}
\begin{equation}
N_{\rm H} \sim  2.8\times10^{21} \cmsq~\left(\frac{\nH}{1\cmcb}\right)^{1/2}  T^{1/2}_{4} \left(\frac{f_{\rm{g}}}{f_{\rm{th}}}\right)^{1/2},
\label{eq:ToTden-col-density}
\end{equation}
where $T_4\equiv T/10^4~{\rm{K}}$, $f_{\rm g}$ is the total mass fraction in gas and $f_{\rm{th}}$ is the fraction of the pressure that is thermal (i.e., $P_{\rm{th}} = f_{\rm{th}} P_{\rm{TOT}}$). In addition, for deriving equation \eqref{eq:ToTden-col-density} we have adopted an adiabatic index, $\gamma = 5/3$, mean particle mass, $\mu = 1.23~{\rm{m_H}}$. After eliminating $N_{\rm H}$ between equation \eqref{eq:Gamma-gasSigma} and equation \eqref{eq:ToTden-col-density}, the photoionization rate can be written as a function of the hydrogen number density
\begin{equation}
\Gamma_{\star} \sim 1.3 \times 10^{-13}~{\rm{s^{-1}}}~\left(\frac{\nH}{1\cmcb}\right)^{0.2} \left(\frac{f_{\rm{g}}}{f_{\rm{th}}}\right)^{0.2} T^{0.2}_4.
\label{eq:Gamma-density}
\end{equation}

Based on equation \eqref{eq:Gamma-density}, the photoionization rate produced by star formation is only weakly sensitive to the temperature, total gas fraction, $f_{\rm g}$, and the fraction of the pressure that is thermal, $f_{\rm{th}}$. The photoionization rate in equation \eqref{eq:Gamma-density} is also very weakly dependent on the gas density. As we will discuss in $\S$\ref{sec:SFR-prof}, our simulations show the same trend and are in excellent agreement with this analytic estimate. However, one should note that due to the simplified assumptions we adopted to derive equation \eqref{eq:Gamma-density}, it provides only an order-of-magnitude estimate for the typical photoionization rate that is produced by young stars in the ISM, on kilo-parsec scales. This relation does not capture the inhomogeneity of the ISM that may result in large fluctuations in the radiation field on smaller scales. We note that the relation we found here between the gas density and stellar photoionization rate is a direct consequence of the underlying star formation law and is independent of the total amount of star formation in a galaxy.

\section{Simulation techniques}
\label{sec:ingredients}

In this section we describe different parts of our simulations. We start with discussing the details of the hydrodynamical simulations that are post-processed with RT calculations. Then we explain our RT that includes the radiation from local stellar radiation, the UVB and recombination radiation.

\subsection{Hydrodynamical simulations}
We use a set of cosmological simulations that are performed using a modified and extended version of the smoothed particle hydrodynamics (SPH) code \Gadget (last described in \citealp{Springel05}). The physical processes that are included in the simulations are identical to what has been used in the reference simulation of the Overwhelmingly Large Simulations (OWLS) described in \citet{Schaye10}. Briefly, we use a subgrid pressure-dependent star formation prescription of \citet{Schaye08} which reproduces the observed Kennicutt-Schmidt law. We use the chemodynamics model of \citet{Wiersma09b} which follows the abundances of eleven elements assuming a \citet{Chabrier03} IMF. These abundances are used for calculating radiative cooling/heating rates, element-by-element and in the presence of the uniform cosmic microwave background and the \citet{HM01} UVB model \citep{Wiersma09a}. We note that the \citet{HM01} UVB model has been shown to be consistent with observations of HI \citep{Altay11,Rahmati12} and metal absorption lines \citep{Aguirre08}. We model galactic winds driven by star formation using a kinetic feedback recipe that assumes $40\%$ of the kinetic energy generated by Type II SNe is injected as outflows with initial velocity of $600~{\rm{kms^{-1}}}$ and with a mass loading parameter $\eta = 2$ \citep{DallaVecchia08}.

We adopt cosmological parameters consistent with the most recent WMAP year 7 results: $\{\Om=0.272,\ \Ob=0.0455,\ \Ol=0.728,\ \sigeight=0.81,\ \ns=0.967,\ h=0.704\} $ \citep{Komatsu11}. Our reference simulation has a periodic box of $L = 6.25$ comoving $\Mpch$ and contains $128^3$ dark matter particles with mass $6.3 \times 10^6~\Msunh$ and an equal number of baryons with initial mass $1.4 \times 10^6~\Msunh$. The Plummer equivalent gravitational softening length is set to $\epsilon_{\rm{com}} = 1.95~\kpch$ and is limited to a minimum physical scale of $\epsilon_{\rm{prop}} = 0.5~\kpch$. We also use a simulation with a box size identical to our reference simulation but with 8 (2) times better mass (spatial) resolution to assess the effect of resolution on our findings (see Appendix \ref{ap:res}). 

In our hydrodynamical simulations, ISM gas particles (which all have densities $\nH > 0.1~\cmcb$) follow a polytropic equation of state that defines their temperatures. These temperatures are not physical and only measure the imposed pressure \citep{Schaye08}. Therefore, when calculating recombination and collisional ionization rates, we set the temperature of ISM particles to $T_{\rm{ISM}} = 10^4~\rm{K}$ which is the typical temperature of the warm-neutral phase of the ISM. Furthermore, we simplify our RT calculations by assuming that helium and hydrogen absorb the same amount of ionizing radiation per unit mass and by ignoring dust absorption and the possibility that some hydrogen may be molecular (see \citealp{Rahmati12} for more discussion).

\subsection{Radiative transfer}
For the RT calculations we use \TRAPHIC (see \citealp{Pawlik08,Pawlik11}; \citealp{Rahmati12}). \TRAPHIC is an explicitly photon-conserving RT method designed to exploit the full spatial resolution of SPH simulations by transporting radiation directly on the irregular distribution of SPH particles. 

The RT calculation in \TRAPHIC starts with source particles emitting photon packets to their neighbors. This is done using a set of $N_{\rm{EC}}$ tessellating emission cones, each subtending a solid angle of $4\pi/N_{\rm{EC}}$. The propagation directions of the photon packets are initially parallel to the central axes of the emission cones. In order to improve the angular sampling of the RT, the orientations of these emission cones are randomly rotated between emission time steps. We adopt $N_{\rm{EC}} = 8$ for computational efficiency. However, we note that our results are insensitive to the precise value of $N_{\rm{EC}}$, thanks to the random rotations.  

After emission, photon packets travel along their propagation direction from one SPH particle to its neighbors. Only neighbors that are inside transmission cones can receive photons. Transmission cones are defined as regular cones with opening solid angle $4\pi/N_{\rm{TC}}$, and are centered on the propagation direction. The parameter $N_{\rm{TC}}$ sets the angular resolution of the RT and we adopt $N_{\rm{TC}} = 64$ which produces converged results (see Appendix C1 of \citealp{Rahmati12}). To guarantee the independence of the RT from the distribution of SPH particles, additional virtual particles (ViP) are introduced. This is done wherever a transmission cone contains no neighboring SPH particle. ViPs do not affect the underlying SPH simulation and are deleted after the photon packets are transferred. 

Furthermore, arriving photon packets are merged into a discrete number of reception cones, $N_{\rm{RC}} = 8$. This makes the computational cost of RT calculations with \TRAPHIC independent of the number of radiation sources. This feature is particularly important for the purpose of the present work as it enables the RT calculations in cosmological density fields with large numbers of sources.

Reception cones are also used for emission from gas particles (e.g., ionizing radiation from star-forming gas particles, recombination radiation). This reduces computational expenses while yielding accurate results. Photon packets are isotropically emitted into reception cones which are already in place at each SPH particle. This obviates the need for constructing any additional emission cone tessellations. Using this recipe, the emission of radiation by star-forming gas particles is identical to that of the emission of diffuse radiation by recombining gas particles, and is described in more detail in Raicevic et al. (in prep.).

Photon packets emitted by the three main radiation sources we consider in our study (i.e., UVB, RR and stellar radiation) are channeled in separate frequency bins and are not merged with each other. This enables us to compute the contribution of each component to the total photoionization rate. The total amount of the absorbed radiation, summed over all frequency bins, determines the ionization state of the absorbing SPH particles. The time-dependent differential equations that control the evolution of the ionization states of different species (e.g., H, He) are solved using a sub-cycling scheme that allows us to choose the RT time step independent of the photoionization and recombination times.

In our RT calculations we use a set of numerical parameters identical to that used in \citet{Rahmati12}. These parameters produce converged results. In addition to the parameters mentioned above, we use 48 neighbors for SPH particles, 5 neighbors for ViPs and an RT time step of $\Delta t = 1~{\rm{Myr}} ~\left(\frac{4}{1+z}\right)$. Ionizing photons are propagated at the speed of light, inside the simulation box with absorbing boundaries, until the equilibrium solution for the hydrogen neutral fractions is reached. To facilitate this process, the RT calculation is starting from an initial neutral state, except for the gas at low densities (i.e., $n_{\rm{gas}} < 1\times 10^{-3}~\cmcb$) and high temperatures (i.e., $T > 10^5$ K) which is assumed to be in equilibrium with the UVB photoionization rate and its collisional ionization rate. Typically, the average neutral fraction in the simulation box does not evolve after 2-3 light-crossing times (the light-crossing time for the $L_{\rm{box}} = 6.25$ comoving $\Mpch$ is $\approx 7.2$ Myr at z = 3).

We continue this section by briefly describing the implementation of the UVB, diffuse recombination radiation and stellar radiation. 
\begin{figure}
\centerline{\hbox{\includegraphics[width=0.5\textwidth]
             {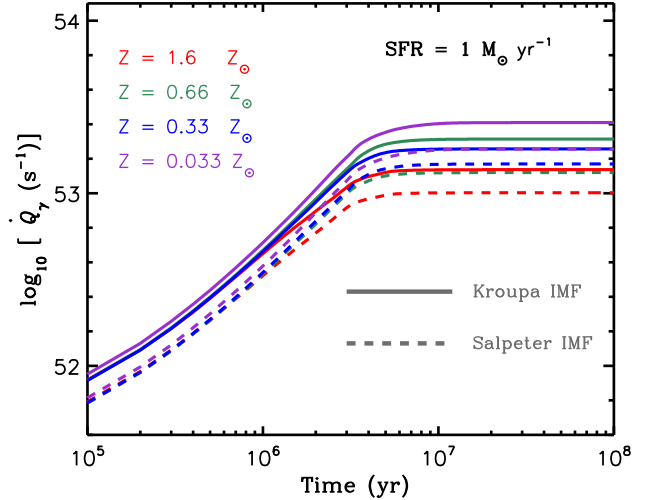}}}
\caption{The number of hydrogen ionizing photons produced for a constant star formation rate of $1~{\rm{M_{\odot}~yr^{-1}}}$ as a function of time since the onset of star formation. These results are calculated using {\small STARBURST99}. Red, green, blue and purple curves indicate metallicities of $ {\rm{Z/Z_{\odot}}} =1.6,~6.6\times10^{-1}~ 3.3\times10^{-1},~\&~ 3.3\times10^{-2}$, respectively. Solid and dashed curves show results for the Kroupa and Salpeter IMF, respectively. All the curves with different metallicities and IMFs converge to similar equilibrium values at $\dot{\mathcal{Q}}_{\gamma} \sim 2 \times 10^{53}$ photons per second, after about $\sim 5-10$ Myr.}
\label{fig:sfr-photons}
\end{figure}

\subsection{Ionizing background radiation and diffuse recombination radiation}
\label{sec:UVB-normalization}
In principle, local sources of ionization inside the simulation box should be able to generate the UVB. However, the box size of our simulations is smaller than the mean free path of ionizing photons which makes the simulated volume too small to generate the observed UVB intensity (see $\S$\ref{sec:SFRmakeUV}). In addition, a considerable fraction of the UVB at $z \lesssim 3$ is produced by quasars which are not included in our simulations. Therefore, we impose an additional UVB in our simulation box.
\begin{figure*}
\centerline{\hbox{\includegraphics[width=0.33\textwidth]
             {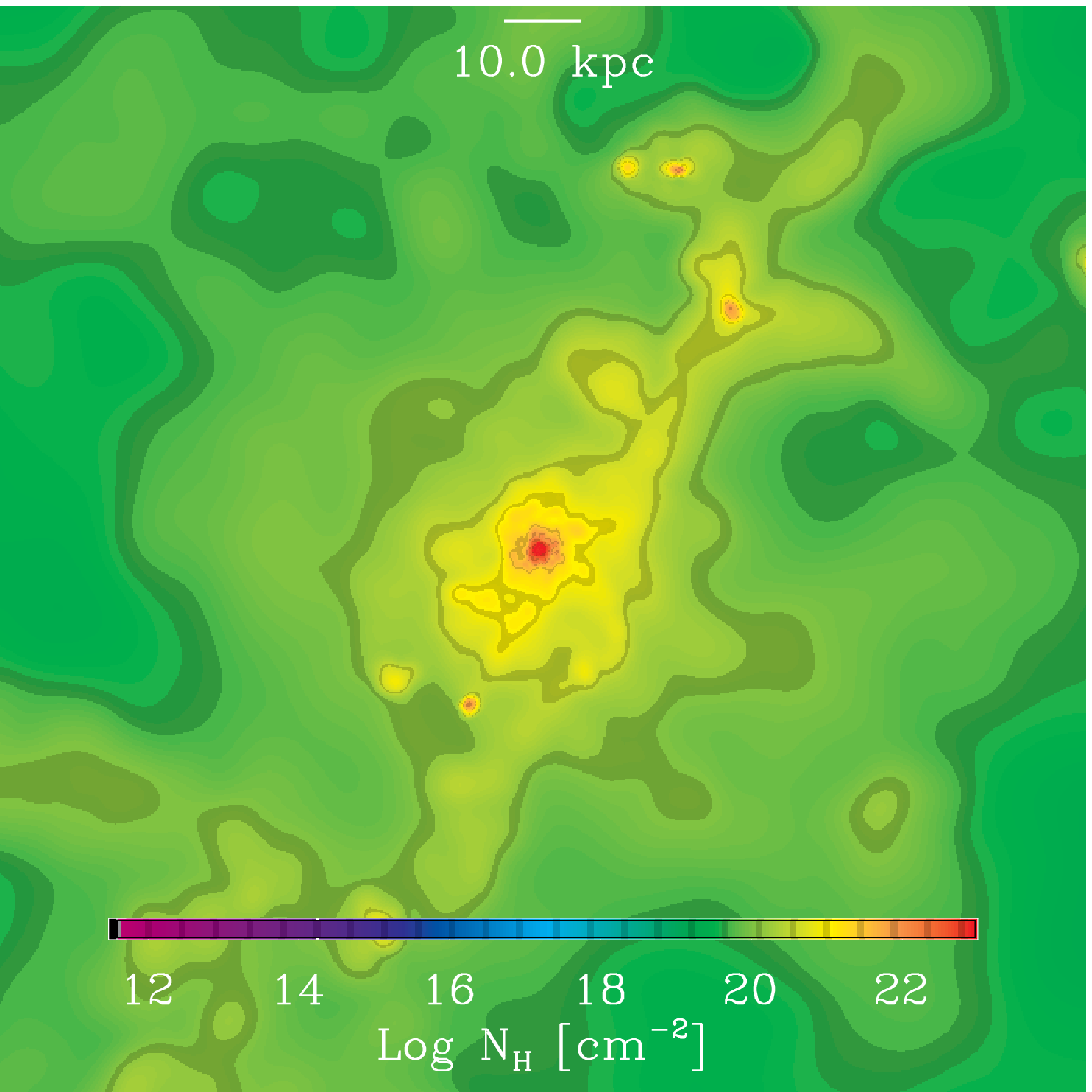}} 
             \hbox{\includegraphics[width=0.33\textwidth]
             {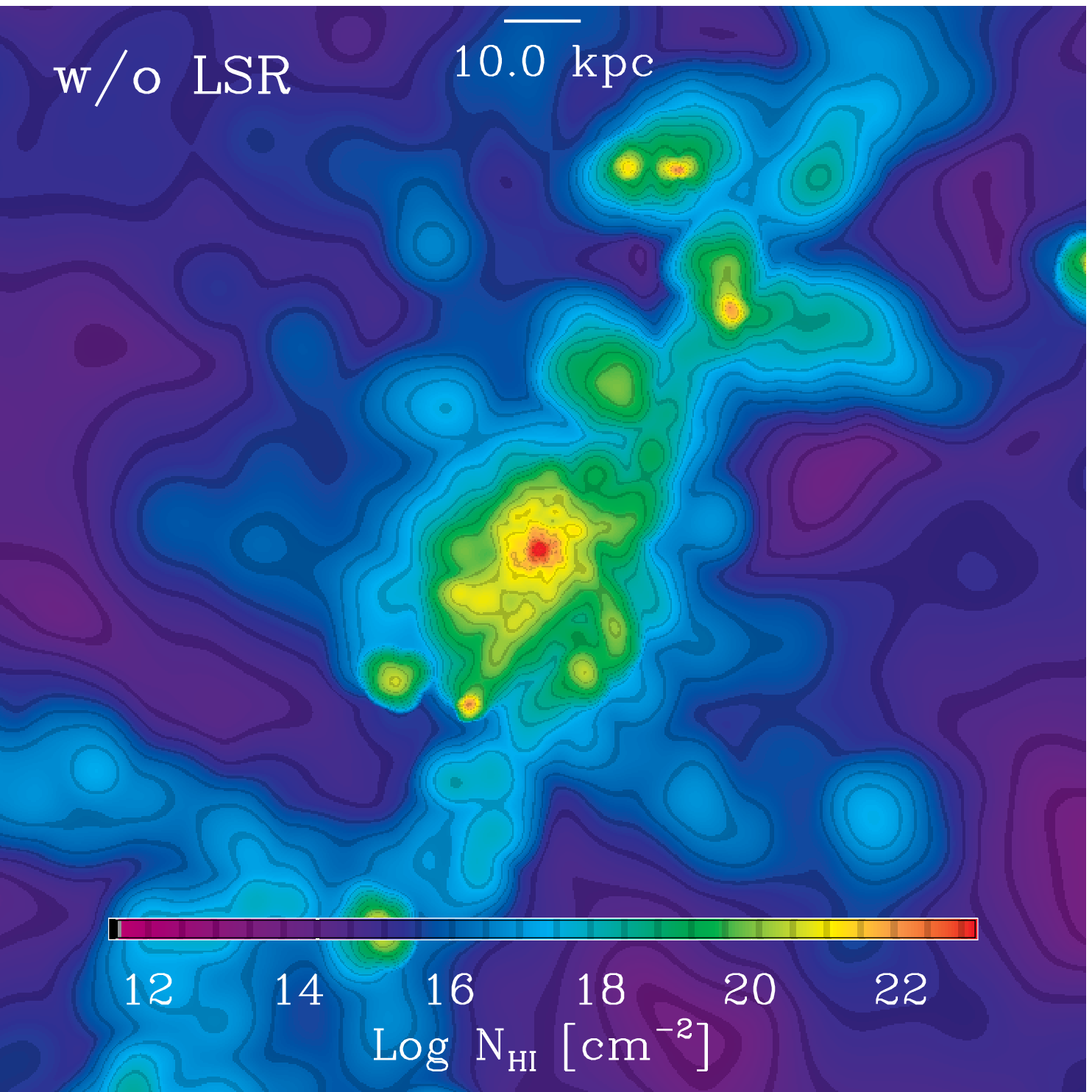}}
             \hbox{\includegraphics[width=0.33\textwidth]
             {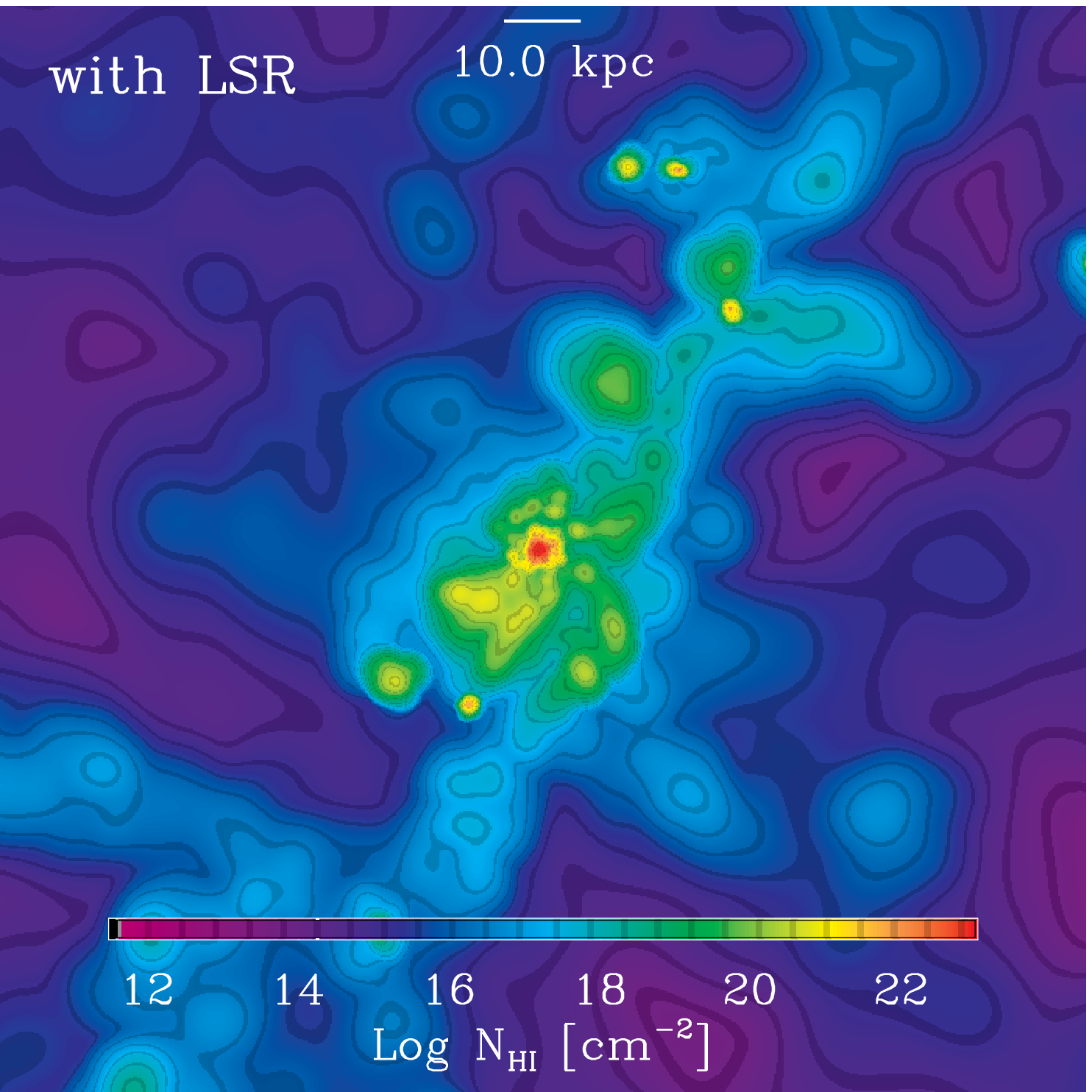}}}
\centerline{\hbox{\includegraphics[width=0.33\textwidth]
             {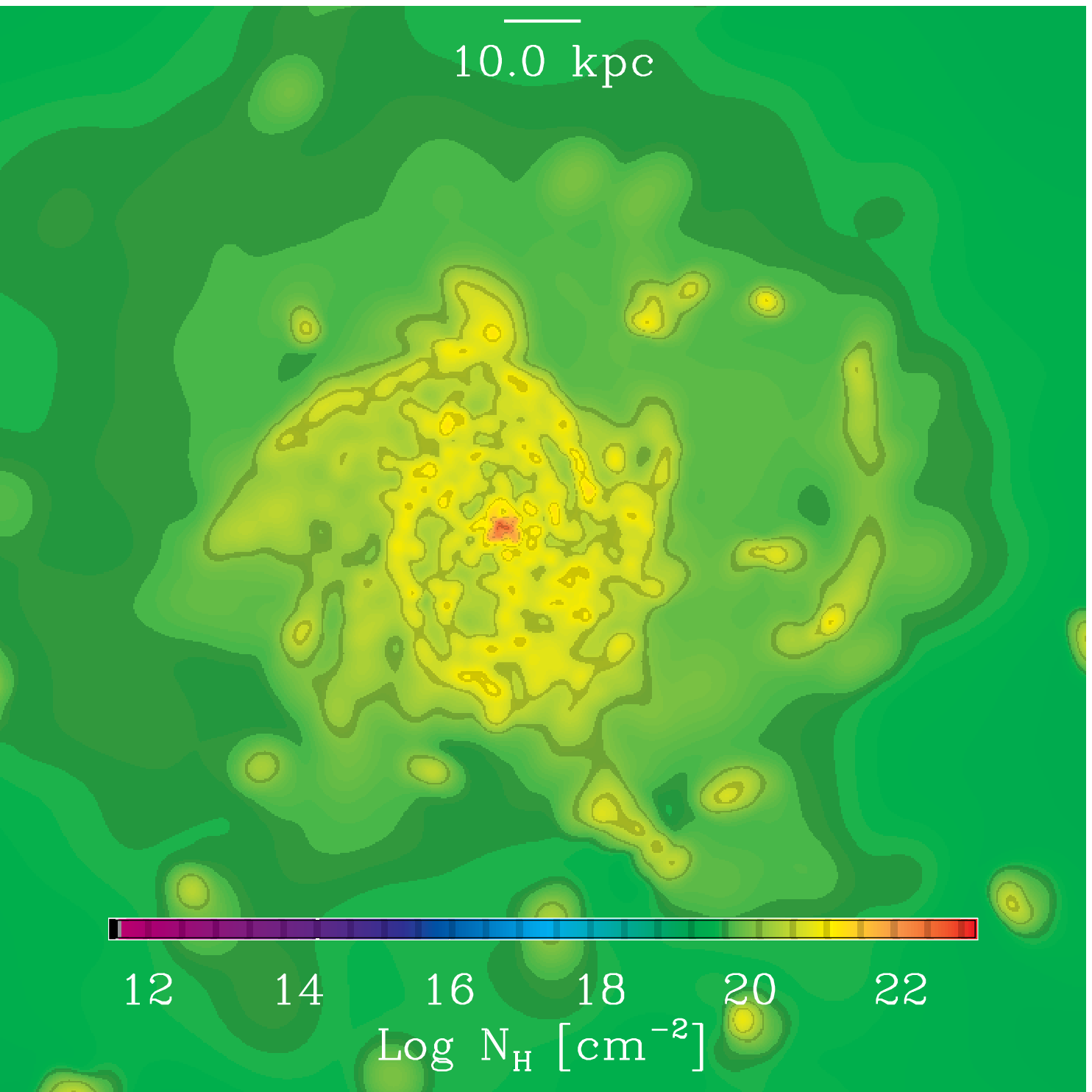}} 
             \hbox{\includegraphics[width=0.33\textwidth]
             {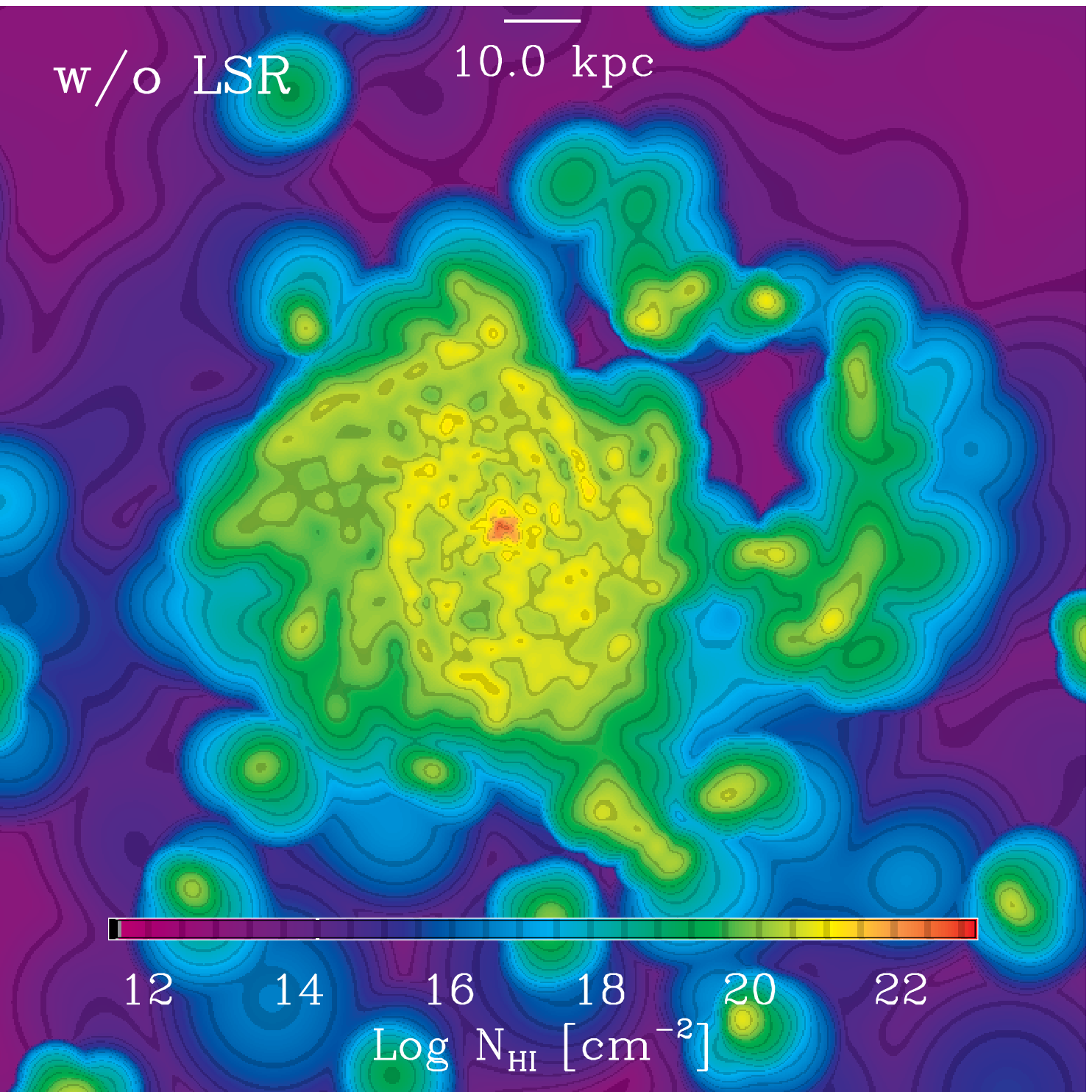}}
             \hbox{\includegraphics[width=0.33\textwidth]
             {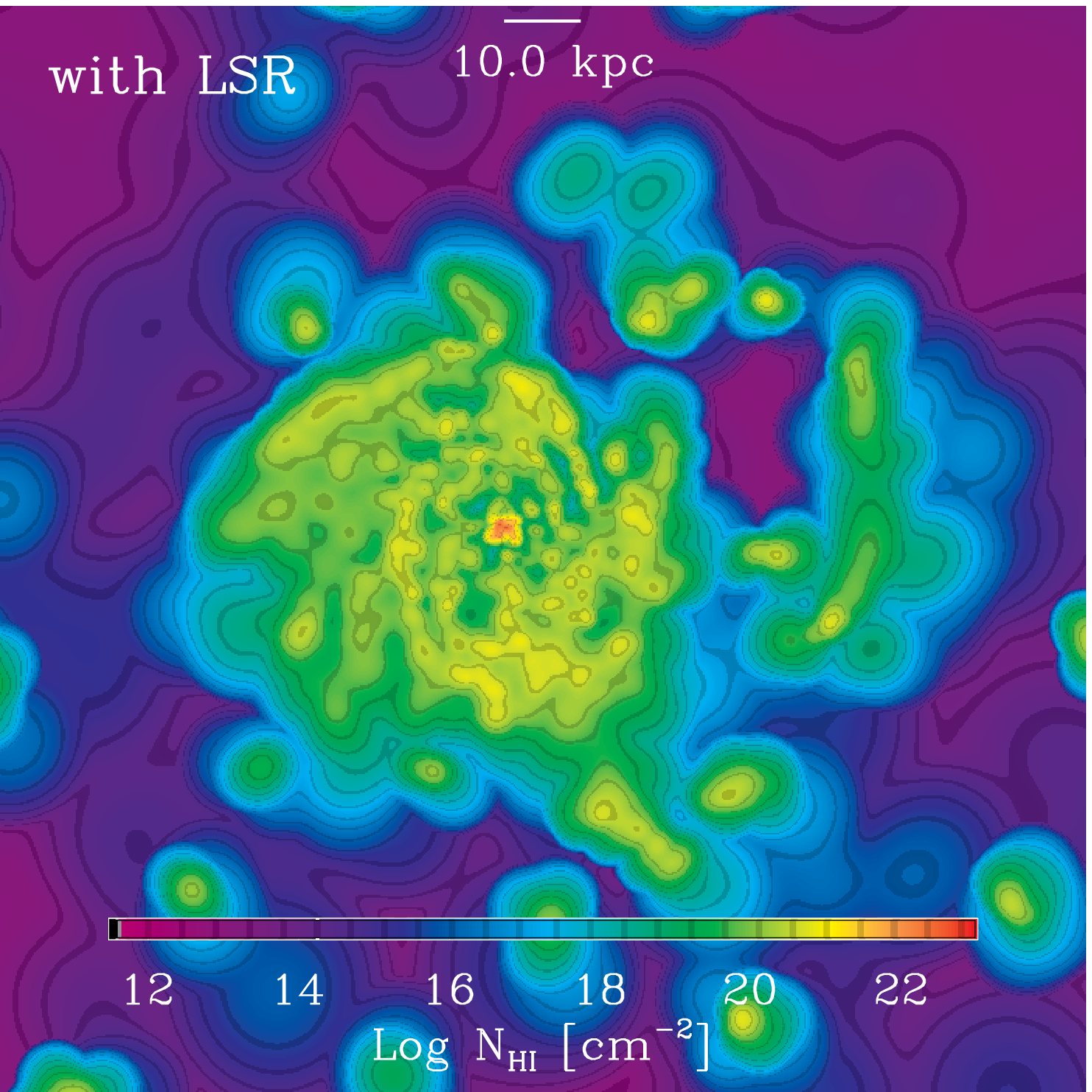}}}
\caption{The HI column density distribution of a Milky Way-like galaxy in the reference simulation at $z=3$ (\emph{top}) and $z = 0$ (\emph{bottom}). The \emph{left panels} show the total hydrogen column density distribution. The \emph{middle panels} show the HI column densities in the presence of collisional ionization, photoionization from the UVB and recombination radiation. In the \emph{right panel}, the HI distribution is shown after photoionization from local stellar radiation is added to the other sources of ionization. The images are $100~\kpch$ (proper) on a side. Comparison between the right and middle panels shows that local stellar radiation substantially changes the HI column density distribution at $\NHI \sim 10^{21}\cmsq$.}
\label{fig:massive-galaxy-z3z0}
\end{figure*}

The implementation of the UVB is identical to that of \citet{Rahmati12}. The ionizing background radiation is simulated as plain-parallel radiation entering the simulation box from its sides and the injection rate of the UVB ionizing photons is normalized to the desired photoionization rate in the absence of any absorption (i.e., the so-called optically thin limit).

We set the effective photoionization rate and spectral shape of the UVB radiation at different redshifts based on the UVB model of \citet{HM01} for quasars and galaxies. The same UVB model has been used to calculate heating/cooling in our hydrodynamical simulations and has been shown to be consistent with observations of HI \citep{Altay11,Rahmati12} and metal absorption lines \citep{Aguirre08} at $z \sim 3$. We adopt the gray approximation instead of an explicit multifrequency treatment for the UVB radiation. Our tests show that a multifrequency treatment of the UVB radiation does not significantly change the resulting hydrogen neutral fractions (see Appendix D1 of \citealp{Rahmati12})

In addition, hydrogen recombination radiation (RR) is simulated by making all SPH particles isotropic radiation sources with emissivities based on their recombination rates. We do not account for the redshifting of the recombination photons and assume that they are monochromatic with energy 13.6 eV (see Raicevic et al. in prep.).

\subsection{Stellar ionizing radiation}
\label{sec:locals-def}

The ionizing photon production rate of star-forming galaxies is dominated by young and massive stars which have relatively short life times. In cosmological simulations with limited mass resolutions, the spatial distribution of newly formed stellar particles (e.g., with ages less than a few tens of Myr) may sample the locally imposed star formation rates relatively poorly. As we show in $\S$\ref{sec:sfVSstar}, the distribution of star formation is better sampled by star-forming SPH particles and this is particularly important for low and intermediate halo masses. For this reason, we use star-forming particles as sources of stellar ionizing radiation. 

For a constant star formation rate, the production rate of ionizing photons reaches an equilibrium value within $\sim 5-10~\rm{Myr}$ (i.e., the typical life-time of massive stars). This equilibrium photon production rate per unit star formation rate can be calculated using stellar population synthesis models. We used {\small STARBURST99} \citep{Leitherer99} to calculate this emissivity for the \citet{Kroupa01} and \citet{Salpeter55} IMF and a metallicity consistent with the typical metallicities of star-forming particles in our simulations (the median metallicity of star-forming particles in our simulations are within the range $10^{-1} \lesssim {\rm{Z/Z}}_{\odot} \lesssim 1$ at redshifts 0 - 5). We found that for a constant star formation rate of $1~\Msun~{\rm{yr}}^{-1}$ and after $\sim 10~\rm{Myr}$, the photon production rate of ionizing photons converges to $\dot{\mathcal{Q}}_{\gamma} \sim 2\times 10^{53}~\rm{s^{-1}}$, which is the value we used in equation \eqref{eq:N-gamma_SFR}. As shown in Figure \ref{fig:sfr-photons}, the ionizing photon production rate varies only by $\lesssim \pm 0.2~\rm{dex}$ if the metallicity changes by $\sim \pm 1~\rm{dex}$. Also, reasonable variations in the IMF (e.g., the Chabrier IMF) do not significantly change our adopted value for the photon production rate. For the spectral shape of the stellar radiation, we adopt a blackbody spectrum with temperature $\rm{T_{bb}} = 5 \times 10^4~\rm{K}$ which is consistent with the spectrum of massive young stars.
\section{Results and discussion}
\label{sec:results}
In this section, we report our findings based on RT simulations that include the UVB, recombination radiation and local stellar radiation. The RT calculations are performed by post-processing a hydrodynamical simulation with $128^3$ SPH particles in a 6.25 comoving $\Mpch$ box and at redshifts $z = 0 - 5$. In $\S$\ref{sec:SFR-prof}, we compare different photoionizing components and assess the impact of local stellar radiation on the HI distribution. In $\S$\ref{sec:sfVSstar} we illustrate the importance of using star-forming SPH particles instead of using young stellar particles as ionizing sources. In $\S$\ref{sec:SFRmakeUV} we show that in our simulations, the predicted intensity of stellar radiation in the IGM is consistent with the observed intensity of the UVB and calculate the implied average escape fraction. In $\S$\ref{sec:fNHI} we show that assumptions about the unresolved properties of the ISM are very important and we discuss the impact of local stellar radiation on the HI column density distribution.

\subsection{The role of local stellar radiation in hydrogen ionization}
\label{sec:SFR-prof}
As we showed in \citet{Rahmati12}, the UVB radiation and collisional ionization are the dominant sources of ionization at low densities where the gas is shock heated and optically thin. However, self-shielding prevents the UVB radiation from penetrating high-density gas. In addition, much of the ISM has temperatures that are too low for collisional ionization to be efficient. Therefore, because UVB photoionization and collisional ionization are both inefficient in those regions, the ionizing radiation from young stars becomes the primary source of ionization. 

Figure \ref{fig:massive-galaxy-z3z0} shows the distribution of neutral hydrogen in and around a galaxy in our reference simulation at redshifts $z = 3$ (top row) and 0 (bottom row) and illustrates that the distribution of HI in galaxies may change significantly as a result of local stellar radiation. The total mass of the halo hosting this galaxy at $z =3$ and 0 is ${\rm{M}_{200}} = 2~\times 10^{11} \Msun$ and $1.1~\times 10^{12} \Msun$, respectively. Comparing the panels in the left and middle columns, we see that while collisional ionization and photoionization by the UVB ionize the low-density gas around the galaxy, they do not change the high HI column densities in the inner regions. Comparing the yellow regions in the right and middle panels of Figure \ref{fig:massive-galaxy-z3z0} demonstrates that local stellar radiation significantly changes the distribution of the HI at $\NHI \sim 10^{21} \cmsq$. 
\begin{figure*}
\centerline{\hbox{\includegraphics[width=0.5\textwidth]
             {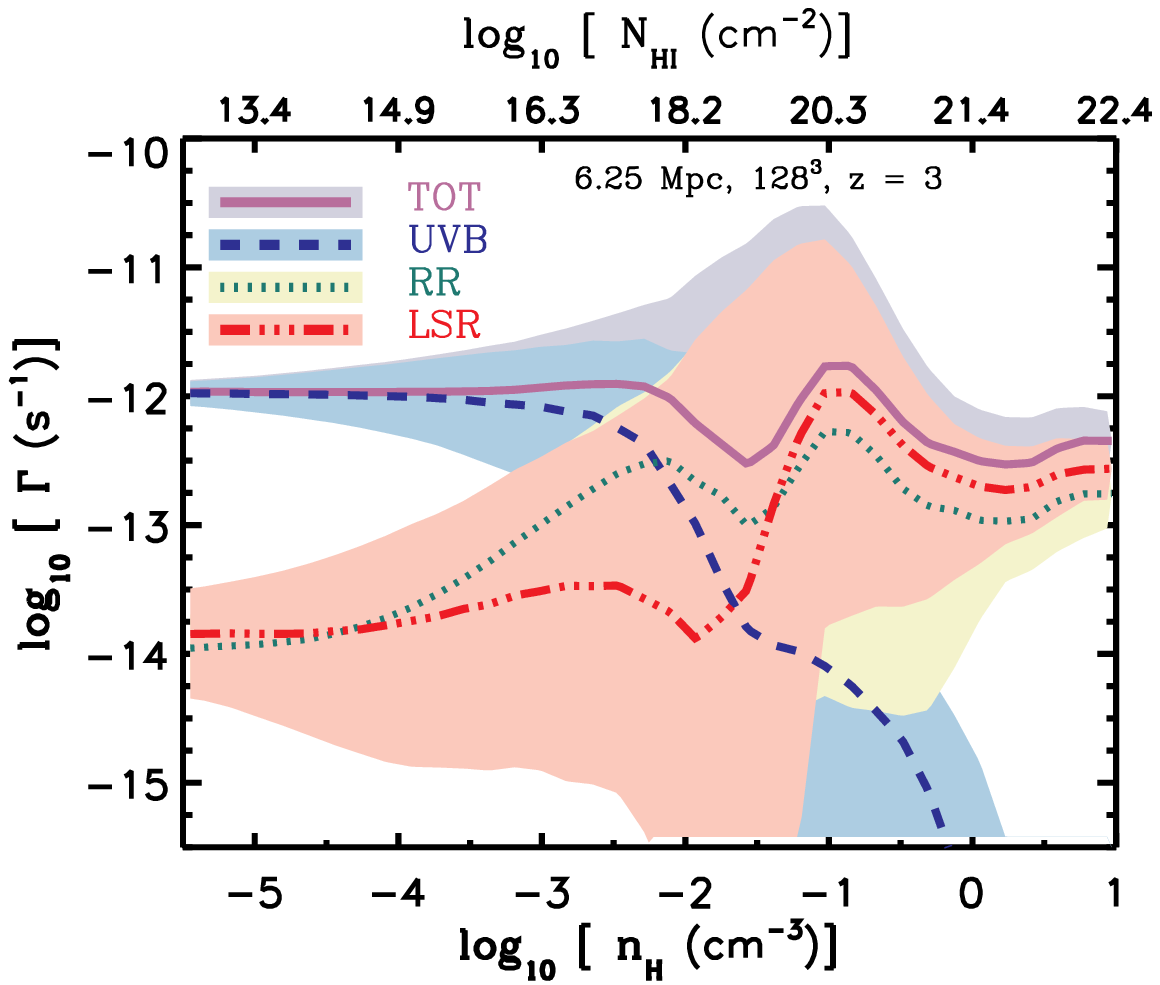}} 
             \hbox{\includegraphics[width=0.5\textwidth]
             {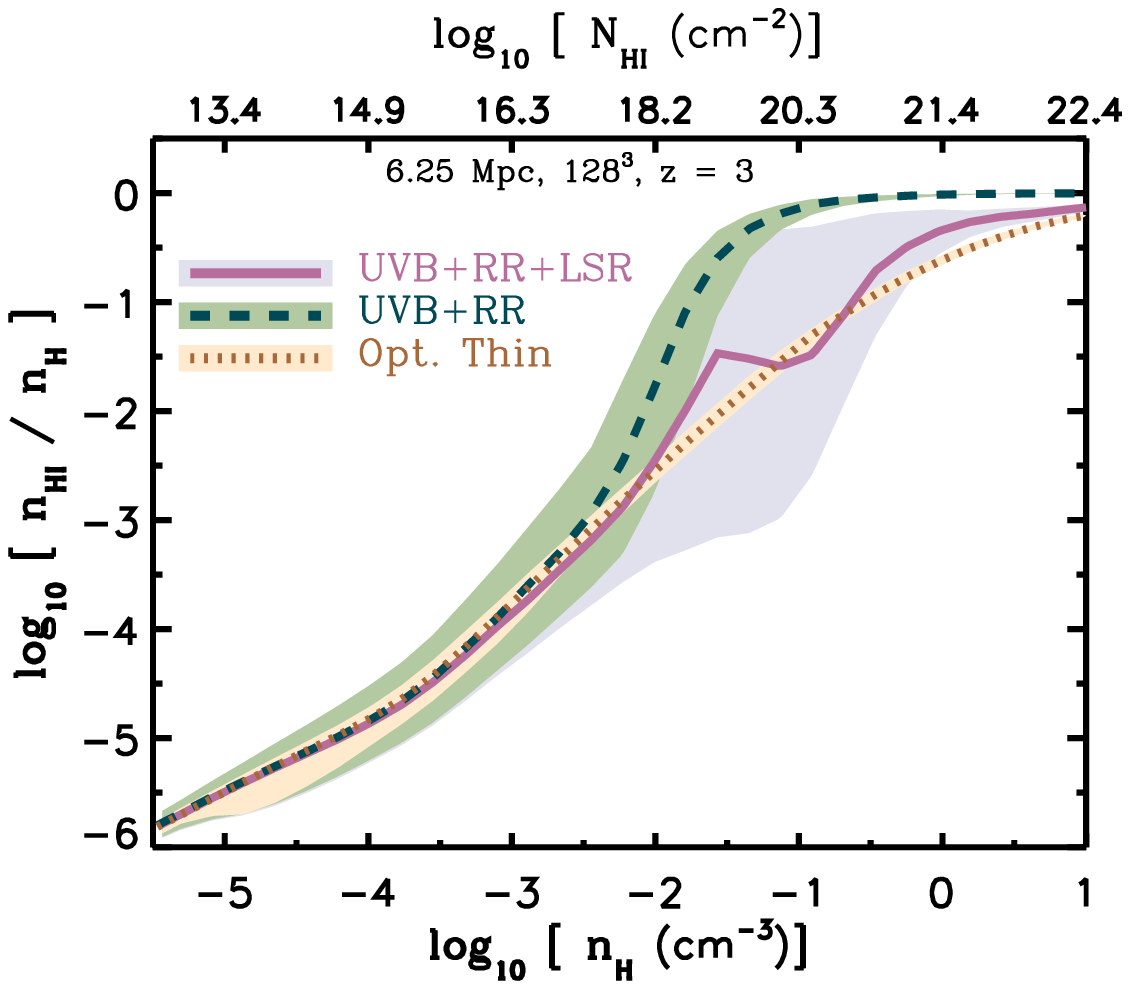}} }
\caption{\emph{Left}: Photoionization rate as a function of density due to different radiation components in the reference simulation at $z =3$. The purple solid curve shows the total photoionization rate. The blue dashed, green dotted and red dot-dashed curves show respectively the photoionization rates due to the UVB, diffuse recombination radiation (RR) and local stellar radiation (LSR). \emph{Right}: The hydrogen neutral fraction as a function of density for the same simulation is shown with the purple solid line. For comparison, the results for a simulation without local stellar radiation (green dashed curve) and a simulation with the UVB radiation in the optically thin limit (i.e., no absorption; brown dotted curve) are also shown. The curves show the medians and the shaded areas around them indicate the $15\%-85\%$ percentiles. HI column densities corresponding to each density in the presence of all ionization sources are shown along the top x-axis. The photoionization due to local stellar radiation exceeds the UVB photoionization rate at high densities and compensates the effect of self-shielding. This produces a hydrogen neutral fraction profile that is very similar to what is expected from the UVB in the optically thin limit.}
\label{fig:Eta_gamma_SFR}
\end{figure*}
\begin{figure*}
\centerline{\hbox{\includegraphics[width=0.5\textwidth]
             {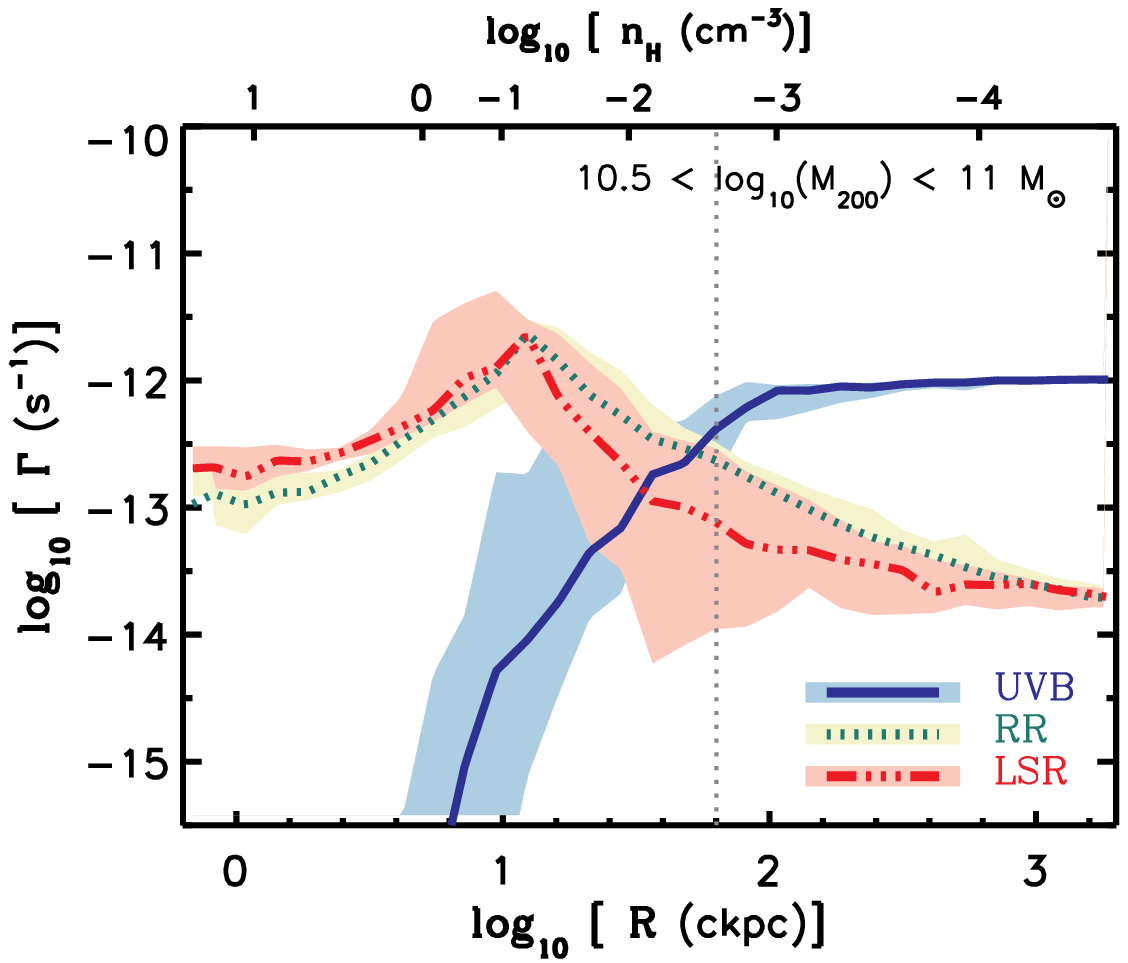}} 
             \hbox{\includegraphics[width=0.5\textwidth]
             {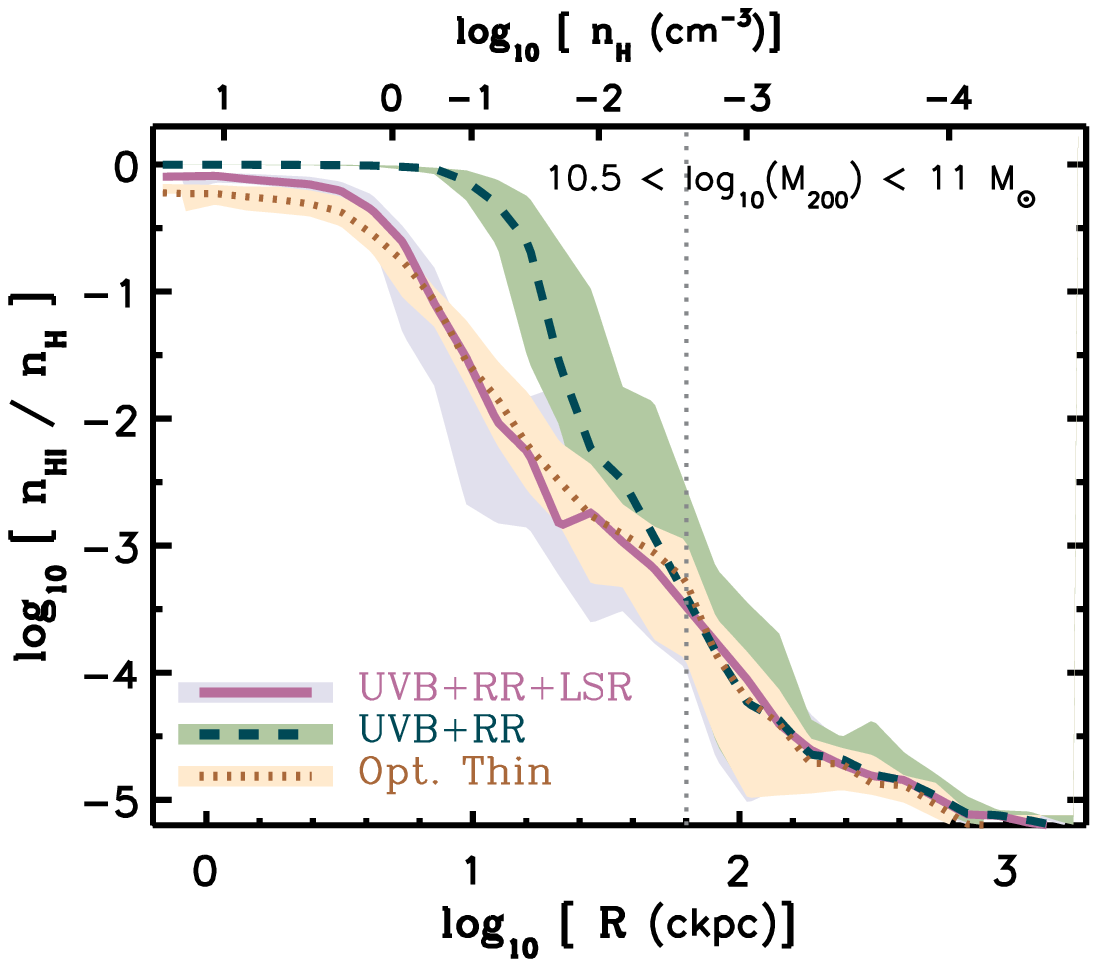}} }
\caption{\emph{Left}: Median photoionization rate profiles (comoving) due to different radiation components for haloes with $10.5 < \log_{10}{\rm{M_{200}}} < 11~\Msun$ at $z = 3$. Blue solid, green dotted and red dot-dashed curves show the photoionization rates of the UVB, recombination radiation (RR) and local stellar radiation (LSR). \emph{Right}: Median hydrogen neutral fraction profiles for the same halos with local stellar radiation (purple solid curve), without local stellar radiation (green dashed curve) and with only the UVB radiation in the optically thin limit (dotted brown). In both panels, the vertical dotted line indicates the median $R_{200}$ radius of the halos in the chosen mass bin. The shaded areas around the medians indicate the $15\%-85\%$ percentiles. The \emph{top axis} in each panel shows the median density at a given comoving distance from the center of the haloes. The photoionization due to local stellar radiation exceeds the UVB photoionization rate close to galaxies and compensates the effect of self-shielding. This produces a hydrogen neutral fraction profile that is very similar to what is expected from the UVB in the optically thin limit.} 
\label{fig:Eta_gamma_SFRHalo}
\end{figure*}

In the left panel of Figure \ref{fig:Eta_gamma_SFR} the contributions of the photoionization rates from different radiation components are plotted against the total hydrogen number density at $z = 3$. The purple solid line shows the total photoionization rate. The blue dashed, green dotted and red dot-dashed lines  show respectively the contribution of the UVB, recombination radiation (RR) and local stellar radiation (LSR). The resulting hydrogen neutral fraction as a function of gas density is shown by the purple solid curve in the right panel of Figure \ref{fig:Eta_gamma_SFR} which illustrates the significant impact of LSR at high densities. We note that there is a sharp feature in the hydrogen neutral fraction at $10^{-1}<\nH<10^{-2}\cmcb$ in the presence of local stellar radiation. This feature corresponds to a sharp drop-off in the photoionization rate from local stellar radiation at the same densities (see the red dot-dashed curve in the left panel of Figure \ref{fig:Eta_gamma_SFR}). The left panel of Figure \ref{fig:Eta_gamma_SFRHalo} shows photoionization rate profiles in spherical shells around haloes with $10.5 < \log_{10}{\rm{M_{200}}} < 11~\Msun$ at $z = 3$. The right panel of Figure \ref{fig:Eta_gamma_SFRHalo} shows the resulting neutral hydrogen fraction profile (purple solid curve) which is compared with the simulation that does not include local stellar radiation (green dashed curve). We note that the trends in the photoionization rate and hydrogen neutral fraction profiles do not strongly depend on the chosen mass bin as long as $ \log_{10}{\rm{M_{200}}} \gtrsim 9.5~\Msun $. 

As Figures \ref{fig:Eta_gamma_SFR} and \ref{fig:Eta_gamma_SFRHalo} show, at low densities (i.e., at large distances from the centers of the halos) the gas is highly ionized by a combination of the UVB and collisional ionization and this optically thin gas does not absorb a significant fraction of the ambient ionizing radiation. Consequently, at $\nH \lesssim 10^{-4} \cmcb$, which corresponds to typical distances ${\rm{R}} \gtrsim 1$ comoving Mpc from the centers of the haloes, the UVB photoionization rate does not change with density. By increasing the density, or equivalently by decreasing the distance to the center of the haloes, the optical depth increases and eventually the gas becomes self-shielded against the UVB. This causes a sharp drop in the UVB photoionization rate at the self-shielding density (see the blue dashed curves in the left panels of Figures \ref{fig:Eta_gamma_SFR} and \ref{fig:Eta_gamma_SFRHalo}). As shown in \cite{Rahmati12}, the UVB photoionization rate shows a similar behavior in the absence of local stellar radiation. However, local stellar radiation increases the photoionization rate at high and intermediate densities. This decreases the hydrogen neutral fractions around the self-shielding densities, allowing the UVB radiation to penetrate to higher densities. Consequently, the local stellar radiation increases the effective self-shielding threshold against the UVB radiation slightly (by $\sim 0.1$ dex) compared to the simulation without the radiation from local stars (not shown).

As shown by the red dot-dashed curve in the left panel of Figure \ref{fig:Eta_gamma_SFR}, at densities $0.1 \lesssim \nH \lesssim 1~\cmcb$ the median photoionization rate due to local stellar radiation increases with decreasing density. The main reason for this is the superposition of radiation from multiple sources (i.e., star-forming SPH particles) as the mean free path of ionizing photons increases with decreasing density (see Figure \ref{fig:mfp}). This is also seen in the left panel of Figure \ref{fig:Eta_gamma_SFRHalo} as increasing stellar photoionization rate with increasing the distance from the center of halos for ${\rm{R}}< 1$ comoving kpc. On the other hand, at densities lower than the star formation threshold (i.e., $\nH < 0.1~\cmcb$), the gas is typically at larger distances from the star-forming regions. Therefore, the photoionization rate of local stellar radiation drops rapidly with decreasing density. The star formation rate density averaged on larger and larger scales becomes increasingly more uniform. This causes the photoionization from galaxies that are emitting ionizing radiation to produce a more uniform photoionization rate at the lowest densities. Note that at the highest densities, the photoionization rate from local stellar radiation agrees well with the analytic estimate presented in $\S$\ref{sec:Gamma-KS}.

In the absence of local stellar radiation, recombination radiation photoionization rate peaks at around the self-shielding density \citep{Rahmati12}. The reason for this is that the production rate of ionizing photons by recombination radiation is proportional to the density of the ionized gas. At densities below the self-shielding threshold, hydrogen is highly ionized and recombination rate is proportional to the total hydrogen density. At much higher densities, the gas is nearly neutral and little recombination radiation is generated. As the left panel in Figure \ref{fig:Eta_gamma_SFR} shows, at high densities the situation changes dramatically if we include local stellar radiation. Since the gas at high densities (e.g., $\nH \gg 10^{-2}\cmcb$) is optically thick, recombination radiation ionizing photons produced at these densities are absorbed locally. In equilibrium, recombination radiation photoionization rate at densities above the self-shielding is therefore a constant fraction of the total ionization rate\footnote{This can be explained noting that in equilibrium, recombination and ionization rates are equal. Depending on temperature, $\approx 40\%$ of recombination photons are ionizing photons which are absorbed on the spot in optically thick gas \cite[e.g.,][]{Osterbrock06}. Therefore, the photoionization rate produced by recombination radiation is also $\approx 40\%$ of the total ionization rate in dense and optically thick gas with $\rm{T} \sim 10^4$ K.}. Below the self-shielding density, the recombination radiation photoionization rate decreases with decreasing density and asymptotes to a background rate. However, in reality recombination photons cannot travel to large cosmological distances without being redshifted to frequencies below the Lyman edge. Therefore, our neglect of the cosmological redshifting of recombination radiation leads us to overestimate the photoionization rate due to recombination radiation on large scales. On the other hand, because of the small size of our simulation box, the total photoionization rate that is produced by recombination radiation remains negligible compared to the UVB photoionization rate and neglecting the redshifting of recombination radiation is not expected to affect our results.

The purple solid curve in the right panel of Figure \ref{fig:Eta_gamma_SFR} shows the hydrogen neutral fractions in the presence of the UVB, recombination radiation and local stellar radiation. For comparison, the hydrogen neutral fractions in the absence of local stellar radiation, and for the optically thin gas that is photoionized only by the UVB, are also shown (with the green dashed and brown dotted curves respectively). Hydrogen at densities $\nH \gtrsim 10^{-1}\cmcb$ is self-shielded and mostly neutral if the UVB and recombination radiation are the only sources of photoionization. However, local stellar radiation significantly ionizes the gas at intermediate and high densities. As mentioned in $\S$\ref{sec:Gamma-KS}, the typical photoionization rate that is produced by star-forming gas is $\Gamma_{\rm{SF}}\sim 10^{-13}~\rm{s^{-1}}$, which is comparable to the photoionization rate of the unattenuated UVB at $z \sim 0$. Consequently, the hydrogen neutral fractions in the presence of local stellar radiation are much closer to the optically thin case. It is also worth noting that the scatter around the median hydrogen neutral fractions is largest for intermediate densities (i.e., $ 10^{-3}\cmcb \lesssim \nH \lesssim 1\cmcb $. This is closely related to the large scatter in the photoionization rate produced by local stellar radiation (see the left panel in Figure \ref{fig:Eta_gamma_SFR}). At these densities, the large scatter in the distances to the nearest sources and RT effects like shadowing produce a large range of photoionization rates. For the gas at very high and very low densities on the other hand, the scatter becomes smaller because the relative distribution of sources with respect to the absorbing gas becomes more uniform. 

The trends discussed so far are qualitatively similar at redshifts other than $z =3$: Although the star formation rates evolve at high densities, the photoionization rate due to local stars is set by the underlying star formation law (see $\S$\ref{sec:Gamma-KS}) which does not change with time. The peak of the stellar photoionization rate at $\nH \sim 10^{-1}~\cmcb$, produced by the superposition of multiple sources, exists at all redshifts but it moves to slightly higher densities at lower redshifts. At densities immediately below the adopted star formation threshold (i.e., $\nH \lesssim 0.1~\cmcb$), the photoionization rate produced by local sources drops rapidly with decreasing density. The distribution of star formation in the simulation box is almost uniform on large scales. Therefore, at the lowest densities, local stellar radiation produces a photoionization rate which is not changing strongly with density. However, at low redshifts (e.g., $z \lesssim 1$), the star formation density decreases significantly and becomes highly non-uniform on the scales probed by our small simulation box. Consequently, the resulting photoionization rate due to young stars does not converge to a constant value at low densities at these redshifts. For the same reason, at low densities the scatter around the median stellar photoionization rate increases with decreasing redshift (not shown). As we will discuss in $\S$\ref{sec:SFRmakeUV}, if we correct for the small size of our simulation box, the asymptotic photoionization rate due to stellar radiation that has reached the IGM, is consistent with the intensity of the observed UVB at the same redshift.
\begin{figure*}
\centerline{\hbox{\includegraphics[width=0.5\textwidth]
             {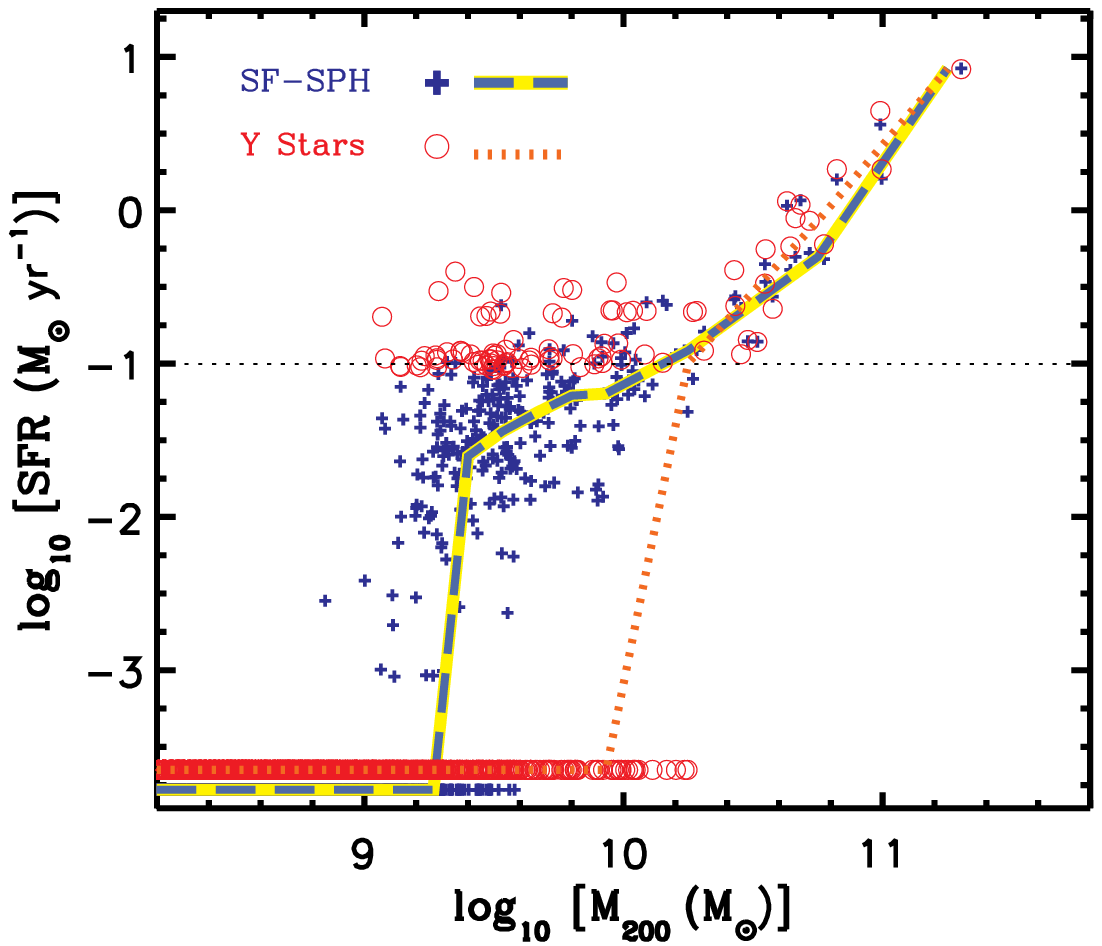}} 
             \hbox{\includegraphics[width=0.5\textwidth]
             {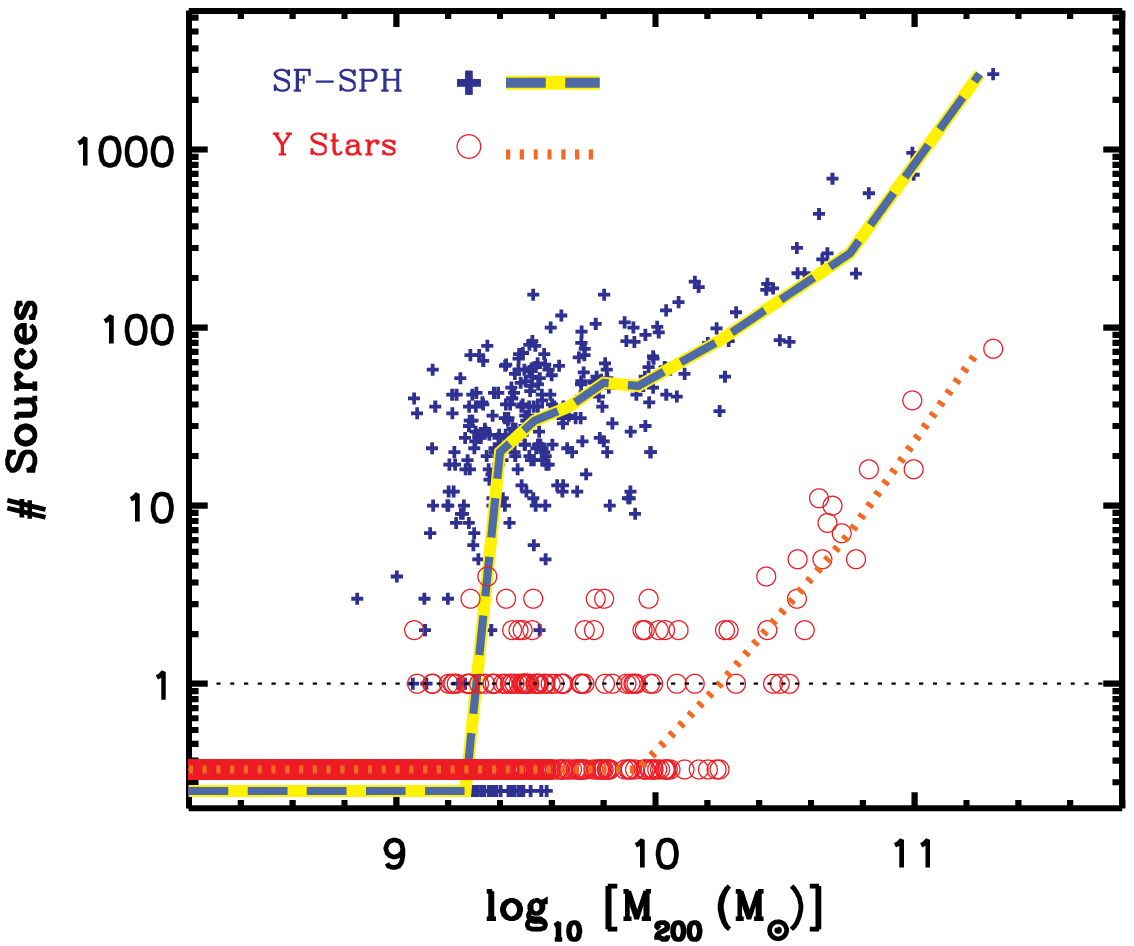}} }
\caption{Star-formation activity of haloes in the reference simulation (6.25 Mpc , $128^3$) at $z=3$. In the \emph{left panel} the blue crosses show the total instantaneous star formation rate for a given halo computed using the gas particles and the red circles show the star formation rate calculated by dividing the total stellar mass formed during the last 20 Myr by 20 Myr. Haloes with zero star formation rates are shown in the bottom of the plot. For massive haloes the two measures agree but as a result of limited mass resolution and the stochastic nature of the star formation algorithm, they start to differ substantially for haloes with $\rm{M_{200}} \lesssim 10^{10}~\Msun$. A large fraction of low-mass haloes does not contain any young stellar particles and the median star formation rate calculated using young stellar particles drops to zero. The \emph{right panel} shows the number of ionizing sources for different halos in the same simulation. While the blue crosses show the number of star-forming gas particles in each halo, red circles indicate the number of young stellar particles (i.e., younger than 20 Myr). In both panels the horizontal dotted lines correspond to a single stellar particle and the blue dashed and red dotted curves show respectively the medians for star-forming gas and young stellar particles.}
\label{fig:source-distribution-MH}
\end{figure*}
\begin{figure*}
\centerline{\hbox{\includegraphics[width=0.5\textwidth]
             {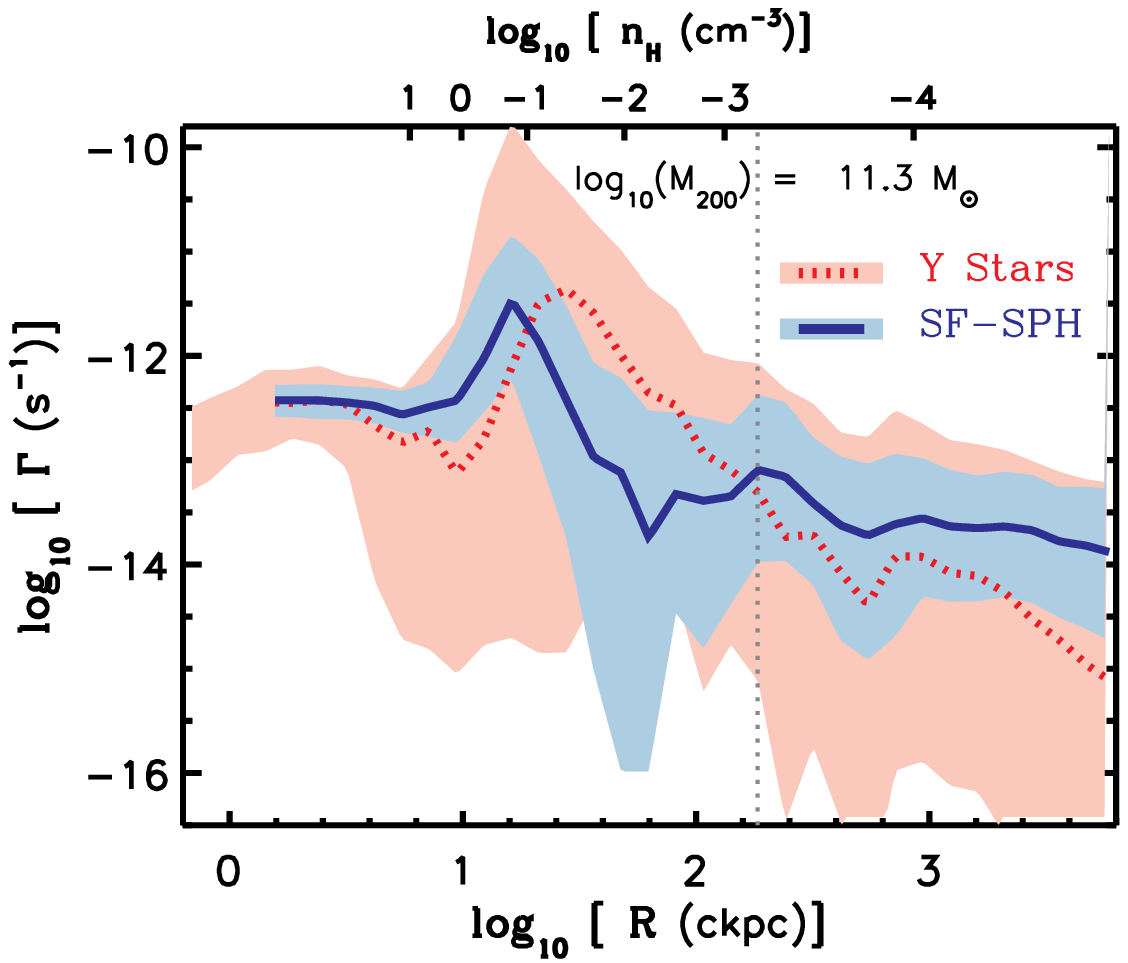}} 
             \hbox{\includegraphics[width=0.5\textwidth]
             {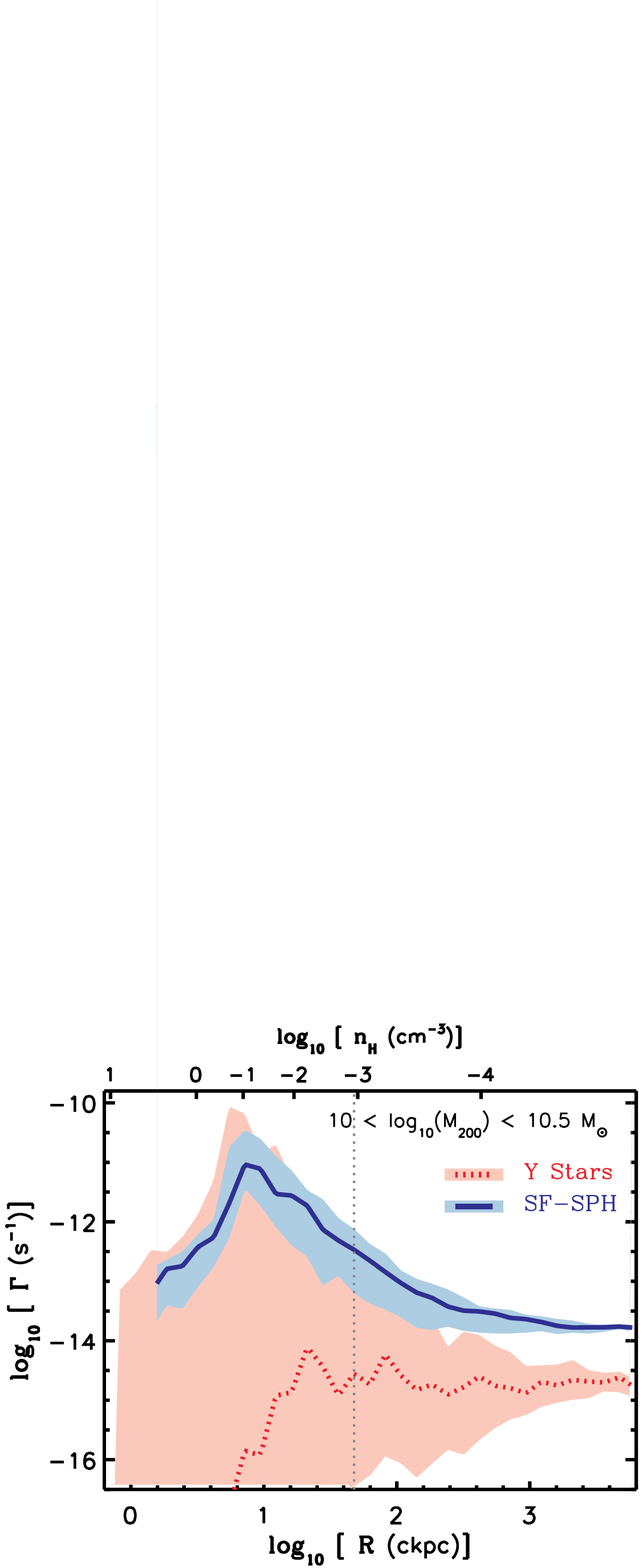}} }
\caption{Median photoionization rate profiles (comoving) for local stellar radiation emitted by star-forming SPH particles (SF-SPH; blue solid curves) and young ($< 20$ Myr) stellar particles (Y Stars; red dotted curves) at $z = 3$. The \emph{left panel} shows the photoionization rate profiles for a halo with ${\rm{M_{200}}}= 2\times10^{11}~\Msun$ while the \emph{right panel} shows the same profiles for haloes with $10 < \log_{10}{\rm{M_{200}}} < 10.5~\Msun$. In both panels the vertical dotted line indicates the median $R_{200}$ radius of the halos in the illustrated mass bin. The shaded areas around the medians indicate the $15\%-85\%$ percentiles. The \emph{top axis} in each panel shows the median density at a given comoving distance from the centers of the haloes. While the photoionization rate profiles produced by star-forming gas particles and young stellar particles are similar for massive haloes (left), they are substantially different for haloes with $\rm{M_{200}} \lesssim 10^{10}~\Msun$ (right).}
\label{fig:Gamma-prof-StarVsSPH}
\end{figure*}
\subsection{Star-forming particles versus stellar particles}
\label{sec:sfVSstar}
In cosmological simulations, the integrated instantaneous star formation rate of the simulation box closely matches the total amount of mass converted into stellar particles. However, due to limited resolution, the spatial distribution of young stellar particles (e.g., those with ages $\sim10$ Myr) may not sample the spatial distribution of star formation in individual haloes very well. 

This issue is illustrated in the left panel of Figure \ref{fig:source-distribution-MH}, where the blue crosses show the total instantaneous star formation rates inside galaxies in our reference simulation at $z =3$. These star formation rates are calculated by the summation of the star formation rates of all SPH particles in a given galaxy. The star formation rates can also be estimated by measuring the average rate by which stellar particles are formed in a given galaxy over some small time interval. The star formation rate calculated using this method (i.e., measuring the rate by which stellar particles are formed during the last 20 Myr), is indicted in the left panel of Figure \ref{fig:source-distribution-MH} by the red circles. For massive galaxies, with high star formation rates, the formation rate of stellar particles agrees reasonably well with the star formation rates computed from the gas distribution. However, most haloes with $\lesssim 10^3$ SPH particles (i.e., $\rm{M_{200}} \lesssim 10^{10}~\Msun$) do not contain any young stellar particles despite having non-zero instantaneous star formation rates. Consequently, for the few low-mass galaxies that by chance contain one or more young stellar particles, the implied star formation rates are much higher than the instantaneous rate that corresponds to the gas distribution. Moreover, as the right panel of Figure \ref{fig:source-distribution-MH} shows, the number of star-forming particles in a given simulated galaxy is $\sim 10^2$ times larger than that of young stellar particles. This ratio could be understood by noting that the observed star formation law implies a gas consumption time scale in the ISM that is $\sim 10^2$ times longer than the life times of massive stars. Hence, if star formation is implemented by the stochastic conversion of gas particles into stellar particles, as is the case here, the number of star-forming (i.e., ISM) particles is expected to be $\sim 10^2$ times larger than the number of young stellar particles. This ratio will be somewhat smaller in starbursts or if gas particles are allowed to spawn multiple stellar particles, but under realistic conditions, star formation will be sampled substantially better by gas particles than by young stellar particles.

As a result of the above mentioned sampling effects, using stellar particles as ionizing sources (as was done in all previous work, e.g., \citealp{Nagamine10,Fumagalli11,Yajima12}) would underestimate the impact of local stellar radiation on the HI distribution for a large fraction of simulated galaxies. We therefore use star-forming SPH particles instead of young stellar particles as local sources of radiation. 

Figure \ref{fig:Gamma-prof-StarVsSPH} illustrates the difference between the photoionization rate profiles produced by star-forming SPH particles (blue solid curves) and young stellar particles (red dotted curves) at $z=3$. The left panel of Figure \ref{fig:Gamma-prof-StarVsSPH} shows this for a halo with ${\rm{M_{200}}} = 2\times 10^{11}~\Msun$ (see the top panels of Figure \ref{fig:massive-galaxy-z3z0}), while the right panel illustrates the results for haloes with $10 < \log_{10}{\rm{M_{200}}} < 10.5~\Msun$. For this comparison we imposed the same total number of emitted photons in the simulation box in both cases. This was done by setting the photon production rates of individual gas particles proportional to their star formation rates (see equation \ref{eq:N-gamma_SFR}) and setting the photon production rate of individual young stellar particles proportional to their masses.

For the massive halo, simulating local stellar radiation using SPH particles results in photoionization rate profiles similar to those produced by using stellar particles. This similarity is mainly due to the agreement between the total star formation rate and the rate by which gas is converted into young stellar particles (see the left panel of Figure \ref{fig:source-distribution-MH}). However, this galaxy has $\sim 50$ times fewer young stellar particles than star-forming SPH particles (see the right panel of Figure \ref{fig:source-distribution-MH}) and as the shaded areas in the left panel of Figure \ref{fig:Gamma-prof-StarVsSPH} show, the scatter in the photoionization rate profile is much larger when stellar particles are used. 

For haloes with slightly lower masses, shown in the right panel of Figure \ref{fig:Gamma-prof-StarVsSPH}, the rates by which gas is converted into young stellar particles are similar to the total star formation rates (see the left panel of Figure \ref{fig:source-distribution-MH}). Despite this agreement, the median photoionization rates for these two cases differ dramatically, being much higher when star-forming gas particles are used as sources.  We conclude that using stellar particles as sources results in the underestimation of the ionization impact of local stellar radiation on most of the galaxies in the simulation. The intensity of the UVB produced by the simulation that uses young stellar particles as sources does not agree with the observed UVB intensity, while using star-forming particles resolves this issue, after correcting for the box size (see $\S$\ref{sec:SFRmakeUV}). 

Although in this section we have shown that using star-forming particles as sources of ionizing radiation helps to resolve the above mentioned sampling issues in post-processing RT simulations, we note that doing the same may not be a good solution for simulations with sufficient resolution to model the cold, interstellar gas phase. Such simulations can capture the effects of the relative motions of stars and gas, e.g., the effect of stars moving out of, or destroying, their parent molecular clouds.
\subsection{Stellar ionizing radiation, its escape fraction and the buildup of the UVB}
\label{sec:SFRmakeUV}
\begin{figure*}
\centerline{\hbox{\includegraphics[width=0.5\textwidth]
             {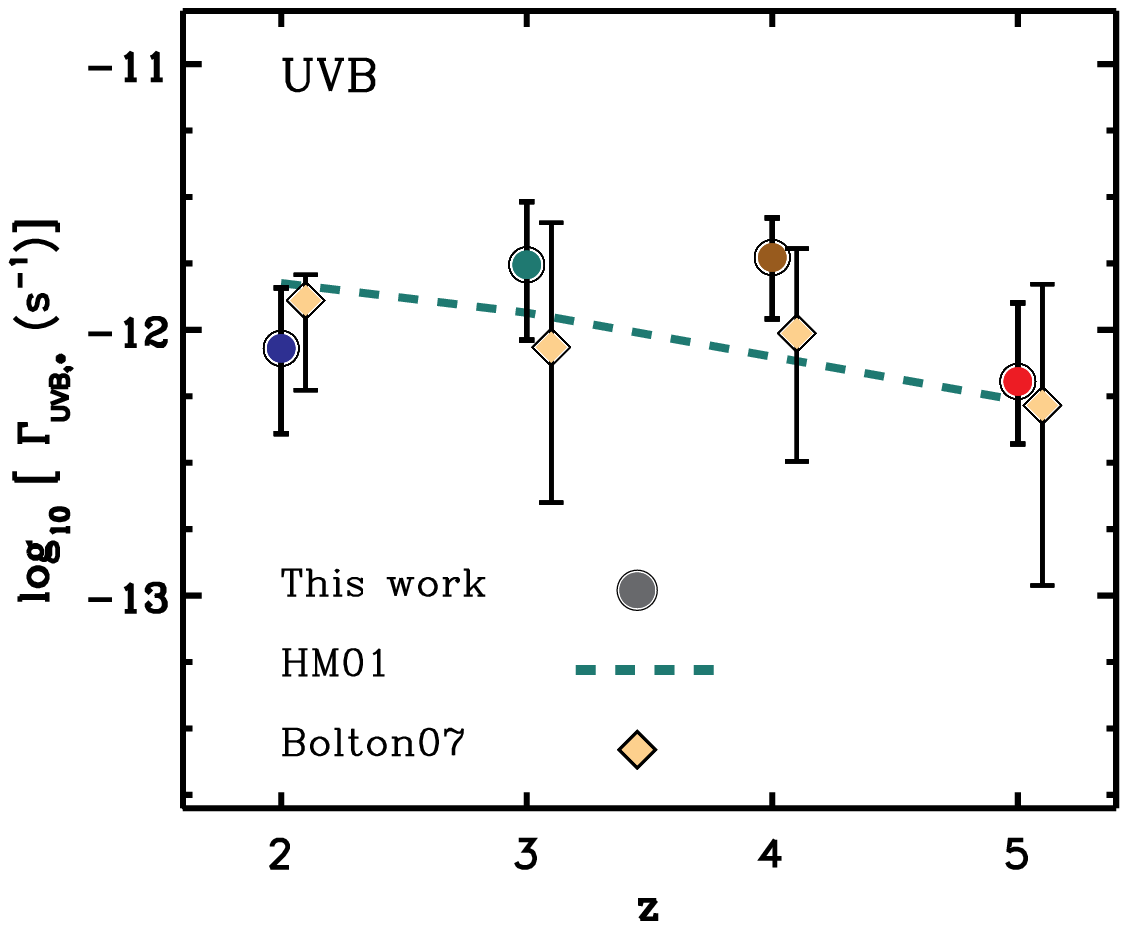}} 
             \hbox{\includegraphics[width=0.5\textwidth]
             {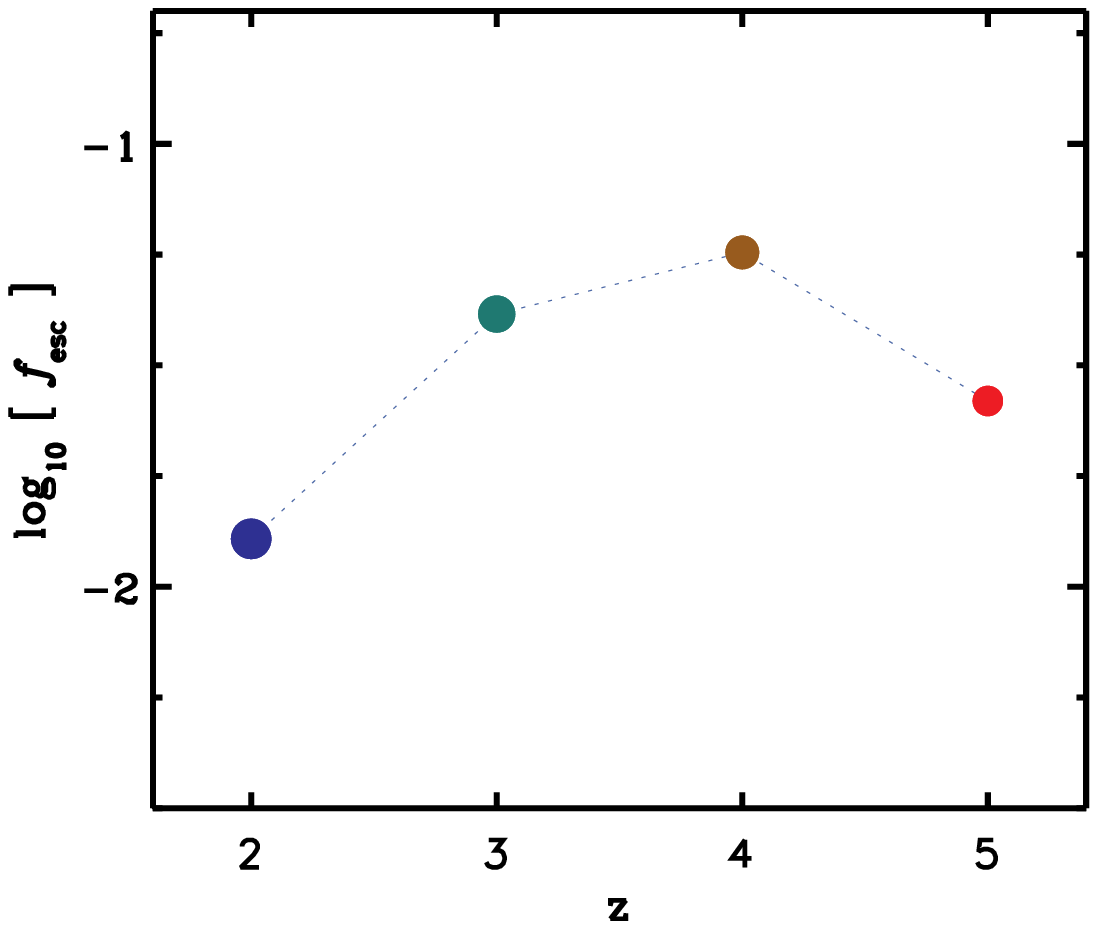}} }
\caption{\emph{Left}: The simulated UVB photoionization rate produced by stellar radiation in our reference simulation is shown with filled circles for different redshifts. The observed mean free paths of ionizing photons have been used to correct for the small size of the simulation box (see text). The error bars represent the $1-\sigma$ errors in these mean free paths. The dashed curve shows the \citet{HM01} UVB photoionization rates that have been used in our simulations as the UVB. The observational measurement of the UVB from the Ly$\alpha$ effective opacity by \citet{Bolton07} is shown using orange diamonds. \emph{Right}: the average escape fraction of stellar ionizing radiation into the IGM calculated based on equation \ref{eq:fesc_SFR_numbers}. Our simulation reproduces the observed UVB photoionization rates between $z = 2$ and 5. The implied average escape fractions are $10^{-2} < f_{\rm{esc}} < 10^{-1}$ between $z=2$ and 5 and they decrease with decreasing redshift below $z=4$ }
\label{fig:sUVB_fesc}
\end{figure*}
Observations show that the luminosity function of bright quasars drops sharply at redshifts $z \gtrsim 2$ \citep[e.g.,][]{Hopkins07} and models for the cosmic ionizing background indicate that star-forming galaxies dominate the production of hydrogen ionizing photons at $z \gtrsim 3$ \citep[e.g.,][]{Haehnelt01,Bolton05,Faucher08}. In the simulations, it should therefore be possible to build up the background radiation from the radiation produced by star-forming galaxies at $z \gtrsim 3$. In this section, we quantify the contribution of the ionizing photons that are produced in stars to the build-up of the UVB radiation in our simulations. In addition, we calculate the implied average escape fraction that is required to generate the observed UVB.

\subsubsection{Generating the UVB}
\label{sec:UVBgen}

\begin{table}
\caption{The comoving mean free path of hydrogen Lyman-limit photons, $\lambda_{\rm{mfp}}$, at different redshifts. From left to right, columns respectively show redshift, $\lambda_{\rm{mfp}}$ in comoving Mpc (cMpc) and the references from which the mean free path values are taken, i.e., B07: \citet{Bolton07}, FG08: \citet{Faucher08}, P09: \citet{Prochaska09} and SC10: \citet{Songaila10}.} 
\begin{center}
\begin{tabular}{lcl}
\hline
Redshift & $\lambda_{\rm{mfp}}$ & Reference \\  
& (cMpc) & \\
\hline 
$z = 2$ & $909\pm 252 $& FG08, SC10 \\
$z = 3$ & $337\pm 170 $& FG08, SC10 \\
$z = 4$ & $170\pm 15.5$& FG08, P09 \\
$z = 5$ & $83.4\pm 21.6$& B07, SC10 \\
\hline
\end{tabular}
\label{tbl:mfp}
\end{center}
\end{table}

Our simulation box is smaller than the typical mean free path of ionizing photons, $\lambda_{\rm{mfp}}$ (see Table \ref{tbl:mfp}). Therefore, the IGM gas can receive stellar ionizing radiation from a region that is larger than the simulation box. We note that using periodic boundaries to propagate the stellar ionizing photons is not a good solution to resolve this issue. In addition to increasing the computational expense, using periodic boundaries for RT would require us to account for the cosmological redshifting of ionizing photons. Moreover, because of the small size of our box, stellar photons traveling along paths that are nearly parallel to a side of the box may never intersect an optically thick absorber, if periodic boundaries are being used for RT. However, we can correct for the small size of the simulation box by requiring consistency between the simulated UV intensities that local sources produce in the IGM and existing observations/models of the UVB. In order to do that, we assume an isotropic and homogeneous universe that is in photoionization equilibrium, and $\lambda_{\rm{mfp}} / (1+z) \ll c/H(z)$, so that we can ignore evolution and redshifting during the travel time of the photons. Then, the volume-weighted mean ionizing flux, $F_{\star}$, from stars is given by:
 \begin{equation}
F_{\star} = \int_0^{\infty} u_{\star}~e^{-\frac{r}{\lambda_{\rm{mfp}}}}\,\mathrm{d}r = \bar{u}_{\star}~\lambda_{\rm{mfp}},
\label{eq:flux_box}
\end{equation}
where $\bar{u}_{\star}$ is the photon production rate per unit comoving volume. Moreover, one can express the mean ionizing flux in equation \eqref{eq:flux_box} in terms of the volume-weighted mean ionizing flux produced by source that are inside the simulation box, $F^{\rm{in}}_{\star}$ 
\begin{equation}
F^{\rm{in}}_{\star} = \int_0^{\alpha L_{\rm{box}}} u_{\star}~e^{-\frac{r}{\lambda_{\rm{mfp}}}}\,\mathrm{d}r ,
\label{eq:flux_inbox}
\end{equation}
where $\alpha \lesssim 1$ is a geometrical factor. In the absence of any absorption and for uniform and isotropic distributions of gas and sources, $\alpha$ is set by the average distance between two random points inside a cube. For a cube with unit length this would yield $\alpha =  0.66$ \citep{Robbins78}. Based on the average distance between the low-density gas (e.g., SPH particles with $\nH \lesssim 10^{-5} \cmcb$) and sources (i.e., star-forming particles) in our simulations, we find that the average value of $\alpha$ is $0.66$ which varies mildly with redshift from $\alpha_0 = 0.54$ at $z = 0$ to $\alpha_5=0.79$ at $z = 5$. 

As Table \ref{tbl:mfp} shows, our simulation box is much smaller than the mean free path of ionizing photons. Therefore, we can use $L_{\rm{box}} \ll \lambda_{\rm{mfp}}$ in equation \eqref{eq:flux_inbox} and get:
\begin{equation}
F^{\rm{in}}_{\star} \approx \bar{u}_{\star}~ \alpha L_{\rm{box}},
\label{eq:flux_boxI}
\end{equation}
Using equation \eqref{eq:flux_box} and \eqref{eq:flux_boxI}, the total stellar UVB radiation flux can thus be written as:
\begin{equation}
F_{\star} \approx \frac{\lambda_{\rm{mfp}}}{\alpha ~L_{\rm{box}}}~F^{\rm{in}}_{\star}.
\label{eq:flux-box}
\end{equation}
Therefore, the volume-weighted photoionization rate due to radiation produced by stars in the simulation, $\Gamma^{\rm{in}}_{\star}$, which is close to the median photoionization rate in low-density gas, can be used to calculate the implied UVB photoionization rate after correcting for the small box size of the simulations:
\begin{equation}
\Gamma_{\rm{UVB},\star} \approx  \frac{\lambda_{\rm{mfp}}}{\alpha~L_{\rm{box}}}~\Gamma^{\rm{in}}_{\star}.
\label{eq:Gamma-box}
\end{equation}

As mentioned earlier, our simulations show that the photoionization rate from local stellar radiation approaches a density independent rate at low densities. We take this photoionization rate as $\Gamma^{\rm{in}}_{\star}$. In addition, we use a compilation of available Lyman-limit mean free path measurements at different redshifts from the literature (i.e., from \citealp{Bolton07,Faucher08,Prochaska09,Songaila10}; see Table \ref{tbl:mfp}). After converting the Lyman-limit mean free paths into the typical mean free path of ionizing photons with our assumed stellar spectrum\footnote{Because the effective hydrogen ionization cross section of stellar ionizing photons that we use in this work, $\bar{\sigma}_{\star} = 2.9 \times 10^{-18} \cms$, their typical mean free paths are $\sim 2$ times longer than the mean free path of Lyman-limit photons which have hydrogen ionizing cross sections ${\sigma}_{0} = 6.8 \times 10^{-18} \cms$.}, we use equation \eqref{eq:Gamma-box} to derive the implied UVB photoionization rate. 

Figure \ref{fig:sUVB_fesc} shows the predicted contribution of stellar radiation to the UVB (filled circles). The error bars reflect the quoted error in the mean free path measurements. For comparison, the observational measurement of the UVB from the Ly$\alpha$ effective opacity by \citet{Bolton07} and the modeled \citet{HM01} UVB photoionization rates are also shown by orange diamonds and green dashed curve, respectively. Both the observational measurements and modeled UVB intensities are in good agreement with our simulation results for $z > 2$. However, their UVB intensity is slightly higher than ours at $z = 2$. The reason for this could be the absence of radiation from quasars in our simulations. 

\begin{figure*}
\centerline{\hbox{\includegraphics[width=0.5\textwidth]
             {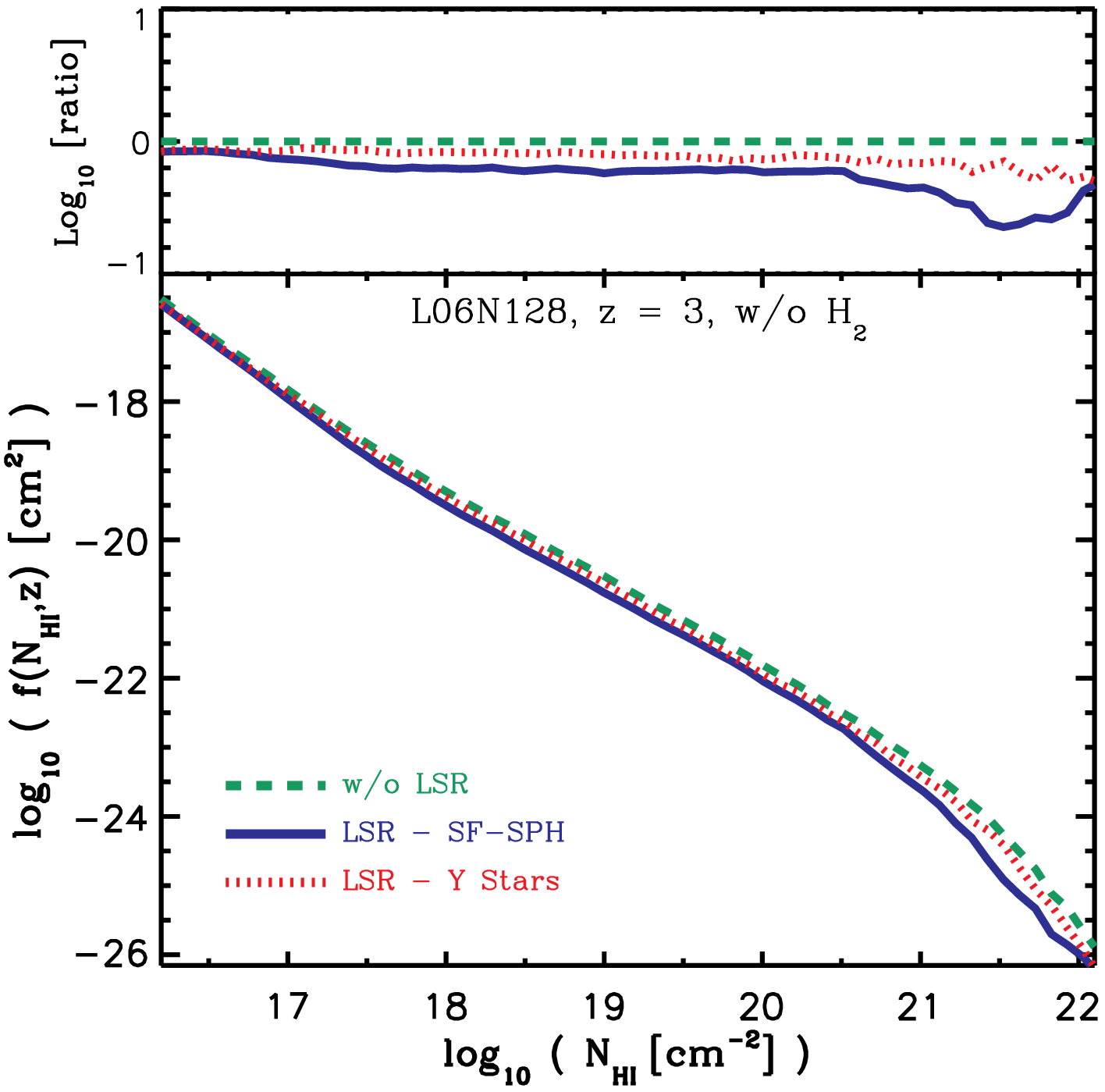}} 
             \hbox{\includegraphics[width=0.5\textwidth]
             {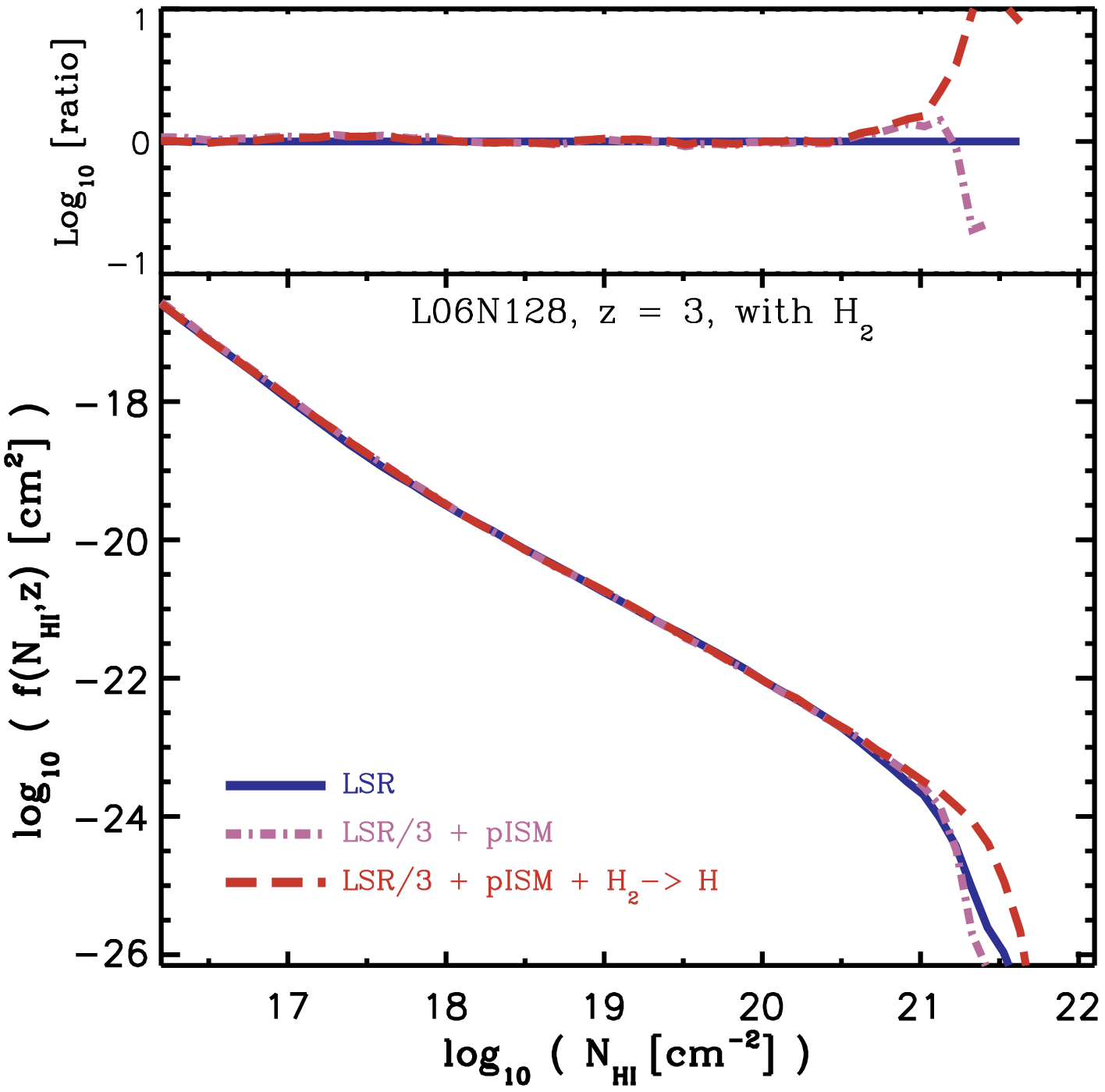}} }
\caption{\emph{Left}: The HI column density distribution function (CDDF) at $z =3$ with different ionizing sources. The blue solid curve shows our reference simulation which includes the UVB, local stellar radiation (LSR) and recombination radiation and the green dashed curve indicates the simulation without local stellar radiation (i.e., with the UVB and recombination radiation). While in the reference simulation star-forming SPH particles are used as ionizing sources, the HI CDDF that is shown with the red dotted curve indicates a simulation in which young stellar particles are used as sources. Using star-forming SPH particles as sources lowers the HI CDDF by $\approx 0.5$ dex for $\NHI \gtrsim 10^{21}\cmsq$. However, using young stellar particles, which results in sampling issues (see $\S$\ref{sec:sfVSstar}), has a weak impact on the HI CDDF. For calculating the HI CDDF, the neutral gas is assumed to be fully atomic (i.e., no $\Hm$). \emph{Right}: The HI CDDF at $z = 3$ in the presence of local stellar radiation (i.e., star-forming SPH particles as sources) for different assumptions about the ISM. The blue solid curve shows our reference simulation in which all H atoms contribute to the absorption but molecular hydrogen does not contribute to the HI CDDF. The orchid dot-dashed curve shows a porous ISM model (see the text) where molecular hydrogen does not absorb ionizing radiation during the RT calculation (it is assumed to have a very small covering fraction). In order to reproduce the observed UVB intensity in this model, the local stellar radiation has been reduced by a factor of 3 compared to the fiducial model. The red long-dashed curve is identical to the orchid dot-dashed curve (porous ISM) but molecular hydrogen is assumed to dissociate into atomic hydrogen before calculating the HI CDDF. At higher HI column densities ($\NHI \gtrsim 10^{21} \cmsq$ ) the HI CDDF is highly sensitive to the assumptions about the unresolved ISM.}
\label{fig:fNHI}
\end{figure*}
\subsubsection{Average escape fractions}
\label{sec:fesc}

The average star formation rate density in our simulations is in good agreement with the observed cosmic star formation rate \citep{Schaye10}. Therefore, the good agreement between the UVB intensities in our simulation and the ones inferred from the observations suggests that at $z \gtrsim 3$ the average escape fractions of stellar ionizing photons in our simulations are also reasonable. However, as we will discuss in Appendix \ref{ap:res}, the structure of the ISM is unresolved in our simulations. Therefore, the fact that the produced escape fractions are reasonable may be coincidental. 

Since in our simulations both the intensity of stellar radiation in the IGM and the photon production rate are known, we can measure the mean star formation rate-weighted escape fraction of ionizing photons from galaxies into the IGM (see Appendix \ref{ap:escape}):
\begin{eqnarray}
 f_{\rm{esc}} &\sim&  10^{-2} ~\left(\frac{\Gamma^{\rm{in}}_{\star}}{10^{-14}~{\rm{s^{-1}}}}\right) \left(\frac{\bar{\sigma}_{\star}}{2.9 \times 10^{-18} \cms}\right)^{-1}   \nonumber \\
   &{}& ~~\times ~~\left(\frac{\dot{\rho}_{\star}}{0.15~\Msun~{\rm{yr^{-1}cMpc^{-3}}}}\right)^{-1} \nonumber \\
   &{}& ~~\times ~~\left(\frac{\alpha}{0.7}\right)^{-1}\left(\frac{ L_{\rm{box}}}{10~{\rm{cMpc}}}\right)^{-1} \left(\frac{1+z}{4}\right)^{-2}.
\label{eq:fesc_SFR_numbers}
\end{eqnarray}
The implied escape fractions are shown in the right panel of Figure \ref{fig:sUVB_fesc}. The simulated escape fractions are $10^{-2} < f_{\rm{esc}} < 10^{-1}$ between $z=2$ and 5 and they decrease with decreasing redshift below $z=4$. This result is consistent with previous observational and theoretical studies \citep[e.g.,][]{Shapley06,Schaye06,Gnedin08,Kuhlen12}.

\subsection{The impact of local stellar radiation on the HI column density distribution}
\label{sec:fNHI}
The observed distribution of neutral hydrogen is often quantified by measuring the distribution of HI absorbers with different strengths in the spectra of background quasars \citep[e.g.,][]{Kim02, Peroux05, Omeara07,Noterdaeme09, Prochaska09, PW09, Omeara12,Noterdaeme12}. The HI column density distribution function (CDDF) is defined as the number of systems at a given column density, per unit column density, per unit absorption length, $d X$:
\begin{eqnarray}
\fNHI \equiv \frac{d^2n}{d \NHI d X}
\equiv \frac{d^2n}{d \NHI d z} \frac{H(z)}{H_0} \frac{1}{(1+z)^2} .
\end{eqnarray}
In order to study the effect of local stellar radiation on the HI CDDF, we project our simulation boxes on a two-dimensional grid and use this to calculate the column densities (see \citealp{Rahmati12} for more details). 

\subsubsection{Local stellar radiation and the HI CDDF at $z =3$}
\label{sec:fNHI-z3}
The simulated HI CDDF at $z =3$ is shown in the left panel of Figure \ref{fig:fNHI} for different ionizing sources. The blue solid curve shows the HI CDDF in our reference simulation which includes local stellar radiation, the UVB and recombination radiation. For comparison, the HI CDDF without local stellar radiation (i.e., only including the UVB and recombination radiation) is shown with the green dashed curve. As the ratio between these two HI CDDFs in the top section of the left panel in Figure \ref{fig:fNHI} illustrates, the effect of local stellar radiation increases with HI column density and reaches a $\sim 0.5 $ dex reduction at $\NHI \gtrsim 10^{21}\cmsq$. For HI column densities lower than $10^{17}\cmsq$ on the other hand, the HI CDDF is insensitive to local stellar radiation. These trends are consistent with previous analytic arguments \citep{Miralda05,Schaye06} and numerical simulations performed by \citet{Fumagalli11}.

As we discussed in $\S$\ref{sec:sfVSstar}, using star-forming SPH particles as ionizing sources (i.e., our reference model) results in a better sampling than using young stellar particles as sources. We showed that if one uses young stellar particles as ionizing sources, the impact of local stellar radiation will be under-estimated in a large fraction of low-mass haloes in our simulations (see Figure \ref{fig:Gamma-prof-StarVsSPH} and \ref{fig:source-distribution-MH}). To illustrate this difference, the HI CDDF for the simulation in which young stellar particles are used, is shown with the red dotted curve in the left panel of Figure \ref{fig:fNHI}. Indeed, the impact of local sources on the HI CDDF is much weaker if we use young stellar particles as ionizing sources. This may partly explain why \citet{Yajima12} found that local sources did not affect the HI CDDF significantly.

\subsubsection{The impact of the unresolved ISM}
\label{sec:ISM-H2}
In our simulations, the ISM is modeled by enforcing a polytropic equation of state on SPH particles with densities $\nH > 10^{-1}\cmcb$. This means that our simulations do not include a cold ($T \ll 10^4$ K) interstellar gas phase. This simple ISM modeling introduces uncertainties in the hydrogen neutral fraction calculations of the dense regions in our simulations. In addition, for gas with $\nH \gg 10^{-2} \cmcb$ the mean free path of ionizing photons is unresolved (see Appendix \ref{sec:ISM-res}).

To estimate the impact of the structure of the ISM on the HI CDDF and the effect of local sources, we introduce a porous ISM model in which we assume that molecular hydrogen is confined to clouds with such small covering fractions that we can ignore them when performing the RT. Following \citet{Rahmati12}, we use the observationally inferred pressure law of \citet{Blitz06} to compute the molecular fraction (see Appendix \ref{ap:H2}), but now we not only subtract the $\Hm$  fraction when projecting the HI distribution to compute the HI CDDF, we also subtract it before doing the RT calculation (note that in the fiducial case we assumed $\Hm$ fractions to be zero during the RT calculation). The conversion of diffuse atomic hydrogen into compact molecular clouds which do not absorb ionizing photons should facilitate the propagation of photons from star-forming regions into the IGM. Indeed, we find that replacing our reference uniform ISM model with the porous ISM model increases the resulting UVB photoionization rate by a factor of $\sim 3$. Therefore, we decreased the photon production rate of local stellar radiation by a factor of 3, such that the model with a porous ISM yields the same UVB intensity that is generated by our reference simulation and is in agreement with the observed UVB. This factor of 3 reduction could be interpreted as reflecting absorptions in molecular clouds that are hosting young stars\footnote{Note that this is not equivalent to $f_{\rm{esc}} = 1/3$. The radiation that leaves star-forming regions is still subject to significant absorption before it reaches the IGM.}. One should note that despite the enforced agreement between the generated UVBs, the stellar photoionization rates at high and intermediate densities are different in the two simulations. 
\begin{figure}
\centerline{\hbox{\includegraphics[width=0.55\textwidth]
             {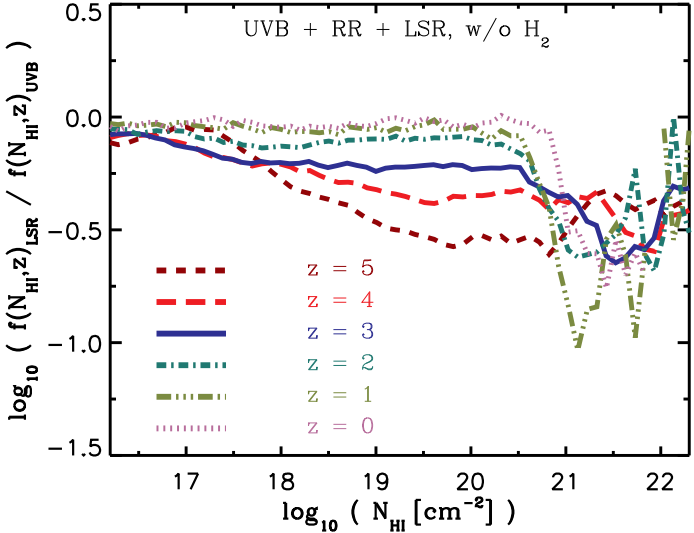}} }
\caption{The ratio between the HI CDDF with and without local stellar radiation at different redshifts. For calculating the HI CDDF, the neutral gas is assumed to be fully atomic (i.e., no $\Hm$). The impact of local stellar radiation decreases the HI CDDF by up to 1 dex. The impact of local stellar radiation on LLSs increases with redshift.}
\label{fig:fNHI-sfr}
\end{figure}

The right panel of Figure \ref{fig:fNHI} shows that for $\NHI < 10^{21} \cmsq$ the HI CDDF in the simulation with a porous ISM (orchid dot-dashed curve) is almost identical to the HI CDDF in our reference simulation (blue solid curve). This suggests that for these column densities the predicted impact of local stellar radiation on the HI CDDF is robust to uncertainties regarding the small scale structure of the ISM. Instead, its impact is controlled by the amount of radiation that is propagating through the properly resolved intermediate densities towards the IGM, which is constrained by the observed UVB intensity. Note, however, that the impact of local sources on the high HI column density part of the distribution (i.e., $\NHI \gtrsim 10^{21}\cmsq$) is in fact highly dependent on the assumptions that are made about the physics of the ISM.

It is also possible that the intense stellar ionizing radiation within the ISM will effectively increase the HI column densities by dissociating hydrogen molecules. However, we implicitly neglected this effect in our simulation with a porous ISM by assuming that molecular clouds are not affected by local stellar radiation. To put an upper limit on the impact of this effect, one can assume that all molecular clouds are completely dissociated by the absorption of stellar radiation. The result of this exercise is shown by the red long-dashed curve in Figure \ref{fig:fNHI}. This shows that accounting for $\Hm$ dissociation could reduce (or even reverse) the impact of local stellar radiation only at the very high column density end of the HI CDDF (i.e., $\NHI \gtrsim 10^{21}\cmsq$). 

We stress that none of our ISM models are realistic. However, by considering very different models, we can nevertheless get an idea of the possible impact of our simplified treatment of the ISM and we conclude that our results are relatively robust for $\NHI \ll 10^{21} \cmsq$.

\subsubsection{Evolution}

The evolution of the impact of local stellar radiation on the HI CDDF is illustrated in Figure \ref{fig:fNHI-sfr} for our fiducial model. Each curve in this figure shows the ratio between the HI CDDF with and without local stellar radiation at a given redshift.  To avoid the uncertainties about the conversion of atomic gas into $\Hm$, the HI CDDFs are computed assuming that the neutral gas is fully atomic (i.e., no $\Hm$). For all redshifts local stellar radiation has only a very small impact on the HI CDDF for $\NHI \ll 10^{17} \cmsq$ but significantly reduces the abundance of systems with $\NHI \gtrsim 10^{21} \cmsq$. The impact of local stellar radiation increases with redshift for $10^{18} < \NHI < 10^{21} \cmsq$. While local sources significantly reduce the HI CDDF for $\NHI \gg 10^{17} \cmsq$ at $z =5$, their effect only becomes significant for $\NHI \gtrsim 10^{21} \cmsq$ at $z =0$. This might be attributed to decrease in the proper sizes of galaxies with redshift.

\begin{figure}
\centerline{\hbox{\includegraphics[width=0.55\textwidth]
             {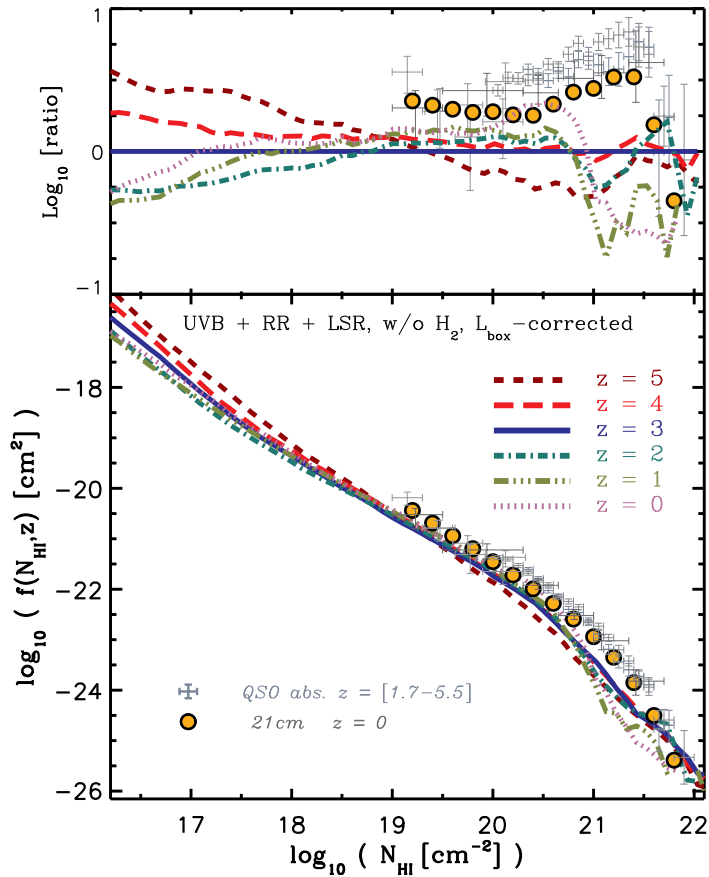}} }
\caption{The HI CDDF at different redshifts. Curves show predictions for the reference simulation in the presence of the UVB, diffuse recombination radiation and local stellar radiation after correcting for the box size (see the text). For calculating the HI CDDF, the neutral gas is assumed to be fully atomic (i.e., no $\Hm$). The observational data points represent a compilation of various quasar absorption line observations at high redshifts (i.e.,  $z = [1.7,5.5]$) taken from \citet{Peroux05} with $z = [1.8,3.5]$, \citet{Omeara07} with $z = [1.7,4.5]$, \citet{Noterdaeme09} with $z = [2.2,5.5]$ and \citet{PW09} with $z = [2.2,5.5]$. The orange filled circles show the best-fit based on the low-redshift 21 cm observations of \citet{Zwaan05}. The \emph{top section} shows the ratio between the HI CDDFs at different redshifts and the predicted CDDF at $z=3$. While simulations without local stellar radiation are in reasonable agreement with the observed HI CDDF (not shown here, but see \citealp{Rahmati12}), in the presence of local stellar radiation the simulated HI CDDFs of LLSs deviate from observations by a factor of $\approx 2$. Local stellar radiation weakens the evolution of the HI CDDFs at $10^{17} \lesssim \NHI \lesssim 10^{20} \cmsq$.}
\label{fig:fNHI-sfr-BCobs}
\end{figure}

In \citet{Rahmati12} we used simulations that include only the UVB and recombination radiation to show that for $10^{18} \lesssim \NHI \lesssim 10^{20} \cmsq$, the HI CDDF does not evolve at $z \le 3$ and increases with increasing redshift for $z > 3$. Figure \ref{fig:fNHI-sfr-BCobs} shows that if we also include local stellar radiation, the weak evolution of the HI CDDF in the Lyman Limit range extends to even higher redshifts (i.e., $z \lesssim 4$). 

The predicted HI CDDFs are compared to observations in Figure \ref{fig:fNHI-sfr-BCobs}. We note that the HI CDDF predicted by our reference simulation is not converged with respect to the size of the simulation box (see Appendix B in \citealp{Rahmati12}). Since the photoionization caused by local stellar radiation only exceeds the UVB photoionization rate close to galaxies (see Figure \ref{fig:Eta_gamma_SFRHalo}), we can assume that the impact of local stellar radiation on the HI CDDF is independent of the size of the simulation box. Therefore, for all the curves shown in Figure \ref{fig:fNHI-sfr-BCobs} we have corrected the box size effect by multiplying the CDDF of the fiducial $L =6.25 \Mpch$ box by the ratio of the CDDF in a converged simulation (with $L =50 \Mpch$) and in our reference simulations with $L =6.25 \Mpch$, both in the absence of local stellar radiation (i.e., with the UVB and recombination radiation). 

As we showed in \citet{Rahmati12}, the simulated HI CDDFs in the presence of the UVB and recombination radiation is in a reasonable agreement with observations (see also \citealp{Altay11}) with a small deviation of $\sim 0.2$ dex. But as the top section of Figure \ref{fig:fNHI-sfr-BCobs} illustrates, the addition of local stellar radiation increases this deviation to $\gtrsim 0.5$ dex. While the difference between the predicted HI CDDFs and observations is larger at the highest HI column densities (i.e., $\NHI \gtrsim 10^{21} \cmsq$), we have seen that in this regime the effect of local sources is sensitive to the complex physics of the ISM, which our simulations do not capture. These very high HI column densities are also sensitive to the strength and the details of different feedback mechanisms (see Altay et al. in prep.). The small but significant discrepancy in the LL and weak DLA regime is therefore more interesting. It is important to confirm this discrepancy with larger simulations, so that a correction for box size will no longer be necessary, but this requires more computing power than is presently available to us.

\section{Discussion and conclusions}
\label{sec:conclusions}
The column density distribution function (CDDF) of neutral hydrogen inferred from observations of quasar absorption lines is the most accurately determined observable of the distribution of gas in the high-redshift Universe. Moreover, the high column density absorbers ($\NHI > 10^{17}\cmsq$) arise in gas that is either already part of the ISM or will soon accrete onto a galaxy \citep{Voort12} and hence these systems directly probe the fuel for star formation. 

To predict the distribution of high column density absorbers, it is necessary to combine cosmological hydrodynamical simulations with accurate radiative transfer (RT) of ionizing radiation. Because of the cost and complexity associated with RT calculations that include many sources, nearly all studies have only considered ionization by the ultraviolet background radiation (UVB), whose intensity can be inferred from observations of the Ly$\alpha$ forest. However, analytic arguments suggest that the radiation field to which high column density absorbers are exposed is typically dominated by local stellar sources \citep{Miralda05,Schaye06}. It is therefore important to investigate whether the remarkable success of simulations that consider only the UVB, such as the agreement with the observed CDDF over 10 orders of magnitude in column density at $z=3$ \citep{Altay11} as well as its evolution down to $z=0$ \citep{Rahmati12}, is compromised when local sources are included. Here we addressed this question by repeating some of the simulations of \citet{Rahmati12} with our RT code TRAPHIC \citep{Pawlik08,Pawlik11,Rahmati12}, but this time including not only the UVB and diffuse recombination radiation, but also local stellar sources.

In agreement with the analytic predictions of \citet{Miralda05} and \citet{Schaye06}, we found that local stellar radiation is unimportant for $\NHI \ll 10^{17} \cmsq$ and dominates over the UVB for high column density absorbers. For all redshifts considered
here (i.e., $z\le 5$), local sources strongly reduce the abundance of systems with $\NHI \gtrsim 10^{21} \cmsq$. The impact of local sources increases with
redshift for $10^{18} < \NHI < 10^{21} \cmsq$. At $z=5$ the CDDF is substantially reduced for $\NHI \gg 10^{17} \cmsq$, but at $z
=0$ the effect only becomes significant for $\NHI \gtrsim 10^{21}
\cmsq$. As a result, the remarkable lack of evolution in the CDDF that we found in \citet{Rahmati12} for $z=0- 3$, and which is also observed, extends to $z=4$ if local sources are taken into account. On the other hand, the agreement with the observed $z\sim 3$ CDDF is not quite as good as before, with the simulations underpredicting the rates of incidence of $10^{19} < \NHI < 10^{21} \cmsq$ absorbers by factors of a few. However, because of the large corrections that we had to make because of the small size of the simulation box used to study the effect of local sources ($6.25 \Mpch$), this discrepancy will have to be confirmed with larger simulations. Moreover, we did not account for possible hydrodynamical effects that might be caused by extra heating due to ionizing radiation from local sources. This process might change the distribution of gas that is affected by local stellar radiation and requires further investigation. 

We found that the average photoionization rate due to young stars in high-density gas is weakly dependent on the gas density and is $\sim 10^{-13} {\rm{s^{-1}}}$. We showed analytically that this rate follows directly from the imposed (and observed) Kennicutt-Schmidt star formation law if we assume that most of the ionizing photons that are produced by star-forming gas are absorbed on scales $\lesssim$ kpc. However, in reality we expect the photoionization rate in the ISM to fluctuate more strongly than predicted by simulations like ours, which lack the resolution required to model the cold, interstellar phase.

Indeed, the spatial resolution that is required for accurate RT of ionizing radiation through the ISM is several orders of magnitude higher than the smallest scales accessible in current cosmological simulations. This makes tackling this problem in the near future hardly feasible and poses a difficult challenge for studying the impact of local stellar radiation on the distribution of HI in and around galaxies. Fortunately, one can circumvent part of the problem by tuning the production rate of ionizing photons (which is equivalent to adjusting the escape fraction of ionizing photons from the unresolved ISM) such that the models reproduce the observed mean photoionization rate in the IGM (after subtracting the contribution from quasars). In other words, if one knows the
amount of ionizing radiation that is required to reach the IGM, its
ionization impact on the intervening gas can be determined even if we cannot predict what fraction escapes from the immediate vicinity of the young stars. 

We adopted this approach but found that tuning was unnecessary for our reference simulation at $z\sim 3$. Moreover, since our simulations also yield star formation histories that are in good agreement with observations (\citealp{Schaye10}), we used them to constrain the implied star formation rate weighted mean escape fraction that relates the predicted star formation rate density to the intensity of the UVB. We found that the average escape fraction in our simulations is $~10^{-2}-10^{-1}$ at $z = 2-5$, which agrees with previous constraints on the escape fraction from observations \citep[e.g.,][]{Shapley06} and theoretical work \citep[e.g.,][]{Schaye06,Gnedin08,Kuhlen12}.

The limited spatial resolution of cosmological simulations mandates the use of simplified models for the structure of the ISM. To estimate the impact of such subgrid models on the CDDF and on the effect of local sources, we varied some of the underlying assumptions. In particular, we considered a porous ISM model which assumes that molecular hydrogen is confined to clouds with such small covering fractions that we can ignore them when performing the RT. We also considered a model which assumes that all molecular clouds are completely dissociated, but not ionised, by the absorption of stellar radiation. Although none of these models are realistic, we used them to estimate the potential impact of our simplified treatment of the ISM. We found that provided that we rescale the source luminosities so that the different models all reproduce the observed background radiation, the models predict nearly the same HI CDDFs in the regime where the absorbers are well resolved. We therefore concluded that our results on the effect of local sources are relatively robust for $\NHI \ll 10^{21} \cmsq$, but that their predicted impact is highly sensitive to the assumptions about the ISM for $\NHI \gtrsim
10^{21}\cmsq$.

Different studies have found qualitatively different results for the impact of local stellar radiation on the CDDF. For instance, \citet{Fumagalli11} used relatively high-resolution zoomed simulations of individual objects to demonstrate that local stellar radiation significantly reduces the abundance of high column density absorbers. On the other hand, some studies using cosmological simulations similar to those presented here (with roughly the same resolution) found that local stellar radiation has a negligible impact on the CDDF \citep{Nagamine10,Yajima12}. 

We found that difference in the resolutions of the simulations that were used in these previous studies may explain their inconsistent findings. Using star particles as sources, as was done by \citep{Nagamine10,Yajima12}, we also found that local sources have a negligible impact on the abundance of strong HI systems. However, we demonstrated that it is possible to dramatically improve the sampling of the distribution of ionizing sources by using star-forming gas particles (i.e., gas with densities at least as high as those typical of the warm ISM), thus effectively increasing the resolution without modifying the time-averaged production rate of ionizing photons. We adopted this strategy in our fiducial models and found, as summarized above, that the radiation from local sources significantly affects the high column density end of the CDDF. This result is in agreement with \citet{Fumagalli11}, \citet{Gnedin10} and analytic estimates of \citet{Miralda05} and \citet{Schaye06}, and confirms that poor sampling of the distribution of ionizing sources can lead to an under-estimation of the impact of local stellar radiation.

Further progress will require higher resolution simulations and, most importantly, more realistic models for the ISM. In the near future it will remain unfeasible to accomplish this in a cosmologically representative volume. Until this challenge is met, predictions for the escape fractions of ionizing radiation averaged over galaxy populations should be considered highly approximate. Predictions for the abundances of LL and weak DLA systems based on models that neglect local sources of stellar radiation should be interpreted with care, particularly for $z > 2$. Predictions for the CDDF in the strong DLA regime ($\NHI \gtrsim 10^{21}\cmsq$) must be considered highly approximate at all redshifts.

\section*{Acknowledgments}
We thank the anonymous referee for a helpful report. We also would like to thank Du\v{s}an Kere\v{s}, J. Xavier Prochaska and Tom Theuns for valuable discussions. The simulations presented here were run on the Cosmology Machine at the Institute for Computational Cosmology in Durham (which is part of the DiRAC Facility jointly funded by STFC, the Large Facilities Capital Fund of BIS, and Durham University) as part of the Virgo Consortium research programme. This work was sponsored with financial support from the Netherlands Organization for Scientific Research (NWO), also through a VIDI grant and an NWO open competition grant. We also benefited from funding from NOVA, from the European Research Council under the European UnionÕs Seventh Framework Programme (FP7/2007-2013) / ERC Grant agreement 278594-GasAroundGalaxies and from the Marie Curie Training Network CosmoComp (PITN-GA-2009-238356). AHP receives funding from the European Union's Seventh Framework Programme (FP7/2007-2013)  under grant agreement number 301096-proFeSsOR.

\appendix
\section{Hydrogen molecular fraction}
\label{ap:H2}
To account for the effect of molecular hydrogen, we adopt the same $\rm{H}_2$ conversion relation that \citet{Altay11} used to successfully reproduce the HI CDDF high-end cut-off. We follow \citet{Blitz06} and adopt an observationally inferred scaling relation between the gas pressure and the ratio between molecular and total hydrogen surface densities:
\begin{equation}
R_{\rm{mol}} = \left(\frac{P_{\rm{ext}}}{P_0}\right)^{\alpha},
\label{eq:Rmol}
\end{equation}
where $R_{\rm{mol}} \equiv \Sigma_{\rm{H_2}} / \Sigma_{\rm{HI}} $, $P_{\rm{ext}}$ is the galactic mid-plane pressure, $\alpha=0.92$ and $P_0/k_b = 3.5 \times 10^4~\rm{cm^{-3}~K}$. Furthermore, if we assume $R_{\rm{mol}}$ gives also the local mass ratio between molecular and atomic hydrogen, the fraction of gas mass which is in molecular form can be written as
\begin{equation}
f_{\rm{H_2}} = \frac{M_{\rm{H_2}}}{M_{\rm{TOT}}} = \frac{M_{\rm{H_2}}}{M_{\rm{H_2}}+M_{\rm{HI}}} = \frac{1}{1+R_{\rm{mol}}^{-1}}.
\label{eq:H2-fraction}
\end{equation}
The last equation also assumes that at high densities ionization is not dominant and the gas is either molecular or neutral. In our simulations we model the multiphase ISM by imposing an effective equation of state with pressure $P \propto \rho^{\gamma_{\rm{eff}}}$ for densities $\rm{n_H > n_H^{\star}}$, where $\rm{n_H^{\star}} = 0.1 ~\rm{cm^{-3}}$ which is normalized to $P_{\star}/k_b = 1.08 \times 10^3 ~\rm{cm^{-3}~K}$ at the threshold. Therefore we have:
\begin{equation}
P_{\rm{ext}} = {P_{\star}} \left(\frac{\nH}{\rm{n_H^{\star}}}\right)^{\gamma_{\rm{eff}}}.
\label{eq:P-nH}
\end{equation}
To make the Jeans mass and the ratio of the Jeans length to the SPH kernel independent of the density, we use $\gamma_{\rm{eff}} = 4/3$ \citep{Schaye08}. Consequently, after combining equations \eqref{eq:Rmol}, \eqref{eq:H2-fraction} \& \eqref{eq:P-nH} the fraction of gas mass which is converted into $\rm{H_2}$, $f_{\rm{H_2}}$ can be written as:
\begin{equation}
f_{\rm{H_2}} = \left(1 + A\left(\frac{\rm{n_H}}{\rm{n_H^{\star}}}\right)^{-\beta}\right)^{-1},
\label{eq:fH2}
\end{equation}
where $A = (P_{\star} / P_0)^{-\alpha}= 24.54$ and $\beta = \alpha~\gamma_{\rm{eff}}= 1.23$. 

\section{Resolution effects}
\subsection{Limited spatial resolution at high densities}
\label{sec:ISM-res}
\begin{figure}
\centerline{\hbox{\includegraphics[width=0.5\textwidth]
             {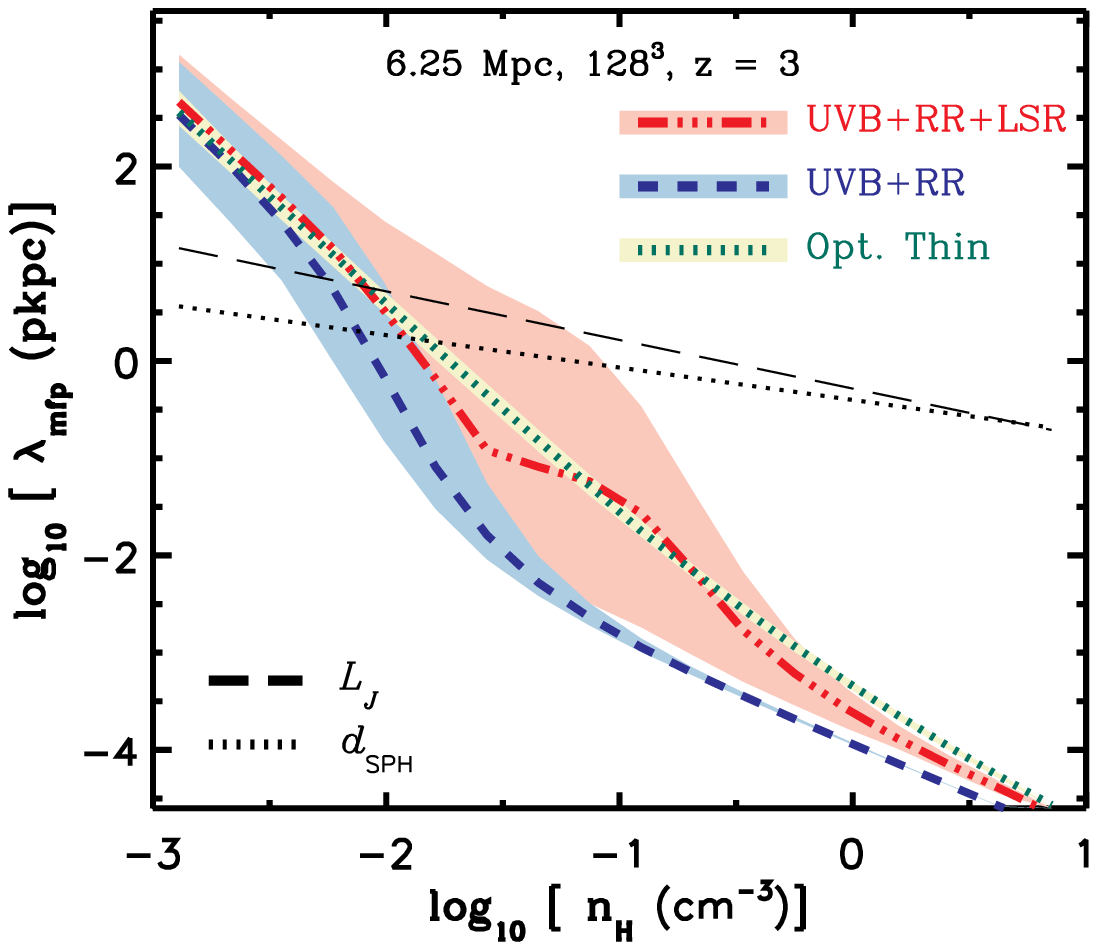}}}
\caption{The mean free path of stellar ionizing photons as a function of hydrogen number density for the reference simulation at $z=3$. The red dot-dashed curve shows the result when all radiation sources, i.e., UVB, local stellar radiation (LSR) and recombination radiation (RR), are included and the blue dashed curve shows the result when local sources of radiation are not present. The green dotted curve shows the mean free path in a simulation that assumes a completely optically thin gas. The typical distance between SPH particles as a function of density is shown by the black dotted line. The Jeans length for a given density is illustrated by the black dashed line. The colored lines indicate the medians and shaded regions around them show the $15\%-85\%$ percentiles.}
\label{fig:mfp}
\end{figure}

At high densities, ionizing photons are typically absorbed within a short distance from the sources. The propagation of ionizing photons in these regions is therefore controlled by distribution of gas on scales that are comparable to the very short mean free path of ionizing photons. In Figure \ref{fig:mfp}, the mean free path of stellar ionizing photons, $\lambda_{\rm{mfp}} = 1/(\nHI~\bar{\sigma}_{\star})$\footnote{$\bar{\sigma}_{\star}$ is the spectrally averaged hydrogen photoionization cross section for stellar ionizing photons. For the spectral shape of the stellar radiation, we adopt a blackbody spectrum with ${\rm{T_{bb}}} = 5\times10^4$ K.}, is plotted as a function of density for our reference simulation at $z = 3$. The blue dashed curve in Figure \ref{fig:mfp} shows the result when only the UVB and recombination radiation are included. The effect of adding local stellar radiation is shown by the red dot-dashed curve. For comparison, also the result for a constant UVB radiation (i.e., optically thin gas) is illustrated with the green dotted curve. The ionization by local stellar radiation decreases the hydrogen neutral fraction at $\nH \gtrsim 10^{-2}\cmcb$ which in turn increases the mean free path of ionizing photons making it comparable to the optically thin case.

The simulations cannot provide any reliable information about the spatial distribution of gas and ionizing sources on scales smaller than the resolution limit (i.e., the typical distance between SPH particles) which is shown by the black dotted line in Figure \ref{fig:mfp}. At densities for which the mean free path of ionizing photons is shorter than the spatial resolution, the RT results may not be accurate since all the photons that are emitted by sources are absorbed by their immediate neighbors. This happens at densities $\nH \gtrsim 10^{-2}\cmcb$ without local stellar radiation (blue dashed curve in Figure \ref{fig:mfp}) and at densities $\nH \gtrsim 10^{-1}\cmcb$ with local stellar radiation (red dot-dashed curve in Figure \ref{fig:mfp}). On the other hand, Rt is irrelevant if the gas is highly neutral, which is predicted to be the case at slightly higher densities (see Figure \ref{fig:Eta_gamma_SFR}).

\subsection{The impact of a higher resolution on the RT}
\label{ap:res}
In \citet{Rahmati12} we showed that the self-shielding limit is not very sensitive to the resolution of the simulation. On the other hand, the small scale structure of the ISM may significantly change the propagation of stellar radiation and their impact on the HI distribution. While this suggests that the impact of local stellar radiation might be sensitive to the resolution of the underlying simulation, our approach of tuning the escaped stellar radiation such that it can generate the desired UVB, circumvent most resolution effects. 

\begin{figure}
\centerline{\hbox{\includegraphics[width=0.5\textwidth]
             {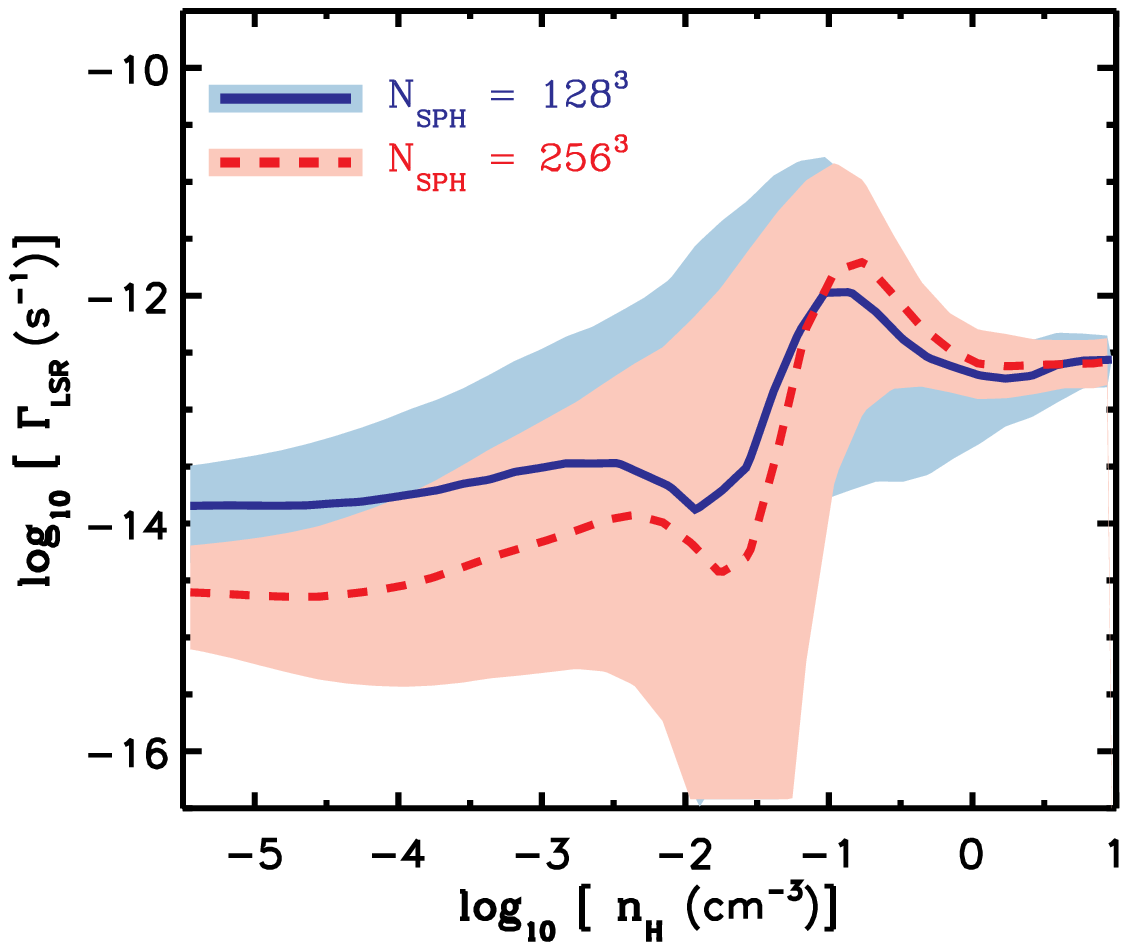}}}
\caption{Comparison between the stellar photoionization rate profiles in the reference simulation and a simulation with 8 times higher resolution at $z =3$. The colored lines indicate the medians and shaded regions around them shows the $15\%-85\%$ distributions.}
\label{fig:ISM-gamma}
\end{figure}

To study the impact of resolution on the propagation of stellar ionizing photons in the ISM and their escape to the IGM, we performed a simulation similar to our reference simulation but using 8 times more dark matter and SPH particles (i.e., using $256^3$ dark matter particles and the same number of SPH particles whose masses are 8 times lower than in our reference simulation). Figure \ref{fig:ISM-gamma} shows that  at $z = 3$, the photoionization rate of stellar radiation in the higher resolution simulation (red dashed curve) is qualitatively similar to that obtained from our reference simulation (blue solid curve). As expected, the stellar photoionization rate at the highest densities is in agreement with our analytic estimate in $\S$\ref{sec:Gamma-KS}. However, because of the shorter inter-particle distances in the higher resolution simulation, the stellar photoionization rate peaks at slightly higher densities. The biggest difference between the stellar photoionization rates of the high-resolution and our reference simulation occurs at the lowest densities. Despite having a $\sim 2$ times higher total star formation rate, the high-resolution simulation results in a UVB which has a $\sim 5$ times lower photoionization rate. This means that the effective escape fraction in the high-resolution simulation is $\sim 10$ times lower than for the reference simulation which has $f_{\rm{esc}} \sim 10^{-2}$. 

Finally, we emphasize that due to our simplified treatment of the ISM, it is not even clear whether the results become more realistic with increasing resolution.

\section{Calculation of the escape fraction}
\label{ap:escape}
Using the equilibrium photon production rate per unit star formation rate from equation \eqref{eq:N-gamma_SFR}, the total production rate of ionizing photons in the simulation box, $\dot{\mathcal{N}}_{\gamma, \star}$, can be calculated as a function of the comoving star formation rate density within the simulation box, $\dot{\rho}_{\star}$, :
\begin{equation}
\dot{\mathcal{N}}_{\gamma, \star} = 2 \times 10^{53}~\left(\frac{\dot{\rho}_{\star}~L_{\rm{box}}^3}{1~\Msun~{\rm{yr^{-1}}}}\right).
\label{eq:phot-rate-density}
\end{equation}
Noting that the typical time the ionizing photons spend inside the simulation box is $\sim \alpha {L_{\rm{box}}}/ (1+z) / c$, and letting $f_{\rm{esc}}$ be the star formation rate-weighted mean fraction of ionizing photons that escape into the IGM, one can calculate the comoving equilibrium photon number density in the IGM:
\begin{equation}
\bar{n}_{\gamma} \sim \frac{f_{\rm{esc}}~\alpha~\dot{\mathcal{N}}_{\gamma, \star}}{L_{\rm{box}}^2 ~c~(1+z)},
\label{eq:escape-number-density}
\end{equation}
where we ignored absorptions outside the host galaxies because $L_{\rm{box}} \ll \lambda_{\rm{mfp}}$. 

The comoving number density of ionizing photons in the simulation box is related to the effective hydrogen photoionization rate they produce in the IGM, $\Gamma^{\rm{in}}_{\star}$:
\begin{equation}
\bar{n}_{\gamma} = \frac{\Gamma^{\rm{in}}_{\star}}{c~\bar{\sigma}_{\star}(1+z)^3},
\label{eq:phot_dens}
\end{equation}
where $\bar{\sigma}_{\star}$ is the effective hydrogen ionization cross section for stellar photons. Combining equations \eqref{eq:escape-number-density} and \eqref{eq:phot_dens} yields the effective escape fraction of ionizing photons from stars into the IGM:
\begin{equation}
 f_{\rm{esc}} \sim \frac{\Gamma^{\rm{in}}_{\star}~L_{\rm{box}}^2 }{\bar{\sigma}_{\star}~\alpha~\dot{\mathcal{N}}_{\gamma, \star}(1+z)^2}.
\label{eq:fesc_SFR}
\end{equation}
After putting numbers in equation \eqref{eq:fesc_SFR} and using equation \eqref{eq:phot-rate-density}, the escape fraction becomes:
\begin{eqnarray}
 f_{\rm{esc}} &\sim&  10^{-2} ~\left(\frac{\Gamma^{\rm{in}}_{\star}}{10^{-14}~{\rm{s^{-1}}}}\right) \left(\frac{\bar{\sigma}_{\star}}{2.9 \times 10^{-18} \cms}\right)^{-1}   \nonumber \\
   &{}& ~~\times ~~\left(\frac{\dot{\rho}_{\star}}{0.15~\Msun~{\rm{yr^{-1}cMpc^{-3}}}}\right)^{-1}    \nonumber \\
   &{}& ~~\times ~~\left(\frac{\alpha}{0.7}\right)^{-1}\left(\frac{ L_{\rm{box}}}{10~{\rm{cMpc}}}\right)^{-1} \left(\frac{1+z}{4}\right)^{-2}.
\label{eq:fesc_SFR_numbersI}
\end{eqnarray}

\end{document}